\documentclass[fleqn,usenatbib, onecolumn]{mnras}

\usepackage{newtxtext,newtxmath}

\usepackage[T1]{fontenc}

\DeclareRobustCommand{\VAN}[3]{#2}
\let\VANthebibliography\thebibliography
\def\thebibliography{\DeclareRobustCommand{\VAN}[3]{##3}\VANthebibliography}

\usepackage{graphicx}	
\usepackage{amsmath}	

\usepackage{physics, bm}
\usepackage{xcolor}

\title{Linear Analyses of Thermal Instability in Stratified Medium}

\author[I. Seno et al. 2025]{
Izumi Seno,$^{1}$\thanks{E-mail: seno.izumi.y0@s.mail.nagoya-u.ac.jp}
Shu-ichiro Inutsuka,$^{1}$
and Jiro Shimoda$^{2}$
\\
$^{1}$Department of Physics, Graduate School of Science, Nagoya University, Furo-cho, Chikusa-ku, Nagoya, Aichi 464-8602, Japan\\
$^{2}$Institute for Cosmic Ray Research, The University of Tokyo, 5-1-5 Kashiwanoha, Kashiwa, Chiba 277-8582, Japan
}

\date{Accepted 2025 December 20. Received 2025 December 20; in original form 2025 October 27}

\pubyear{2025}

\begin{document}
\label{firstpage}
\pagerange{\pageref{firstpage}--\pageref{lastpage}}
\maketitle

\begin{abstract}
Thermal instability in the circum-galactic medium (CGM) can be responsible for the existence of cold clouds (e.g., high-velocity clouds) embedded in a hot diffuse medium (e.g., X-ray emitting gas). 
While many previous studies have analyzed thermal instability in uniform medium, the instability mechanism in gravitationally stratified medium like CGM has not been fully analyzed.
This study investigates how gravity affects the behavior of thermal instability through linear perturbation analyses. 
We find that in stratified medium, thermal instability can drive over-stable modes, a behavior distinctly different from the monotonic growth of thermal instability in a uniform medium. 
Furthermore, we find that the combination of buoyancy and thermal instability drives other two unstable modes. 
Applying our results to a simplified model of the CGM, we estimate the gas accretion rate from the CGM to the Galactic disk and the typical size of high-velocity cloud  driven by thermal instability to be a few $M_\odot {\rm yr}^{-1}$. 
This gas accretion rate is comparable to the observed star formation rate, and hence, the mass in the Galactic disk can be maintained.
Our results provide a theoretical framework for understanding the formation of multi-phase gas, particularly in the CGM.
\end{abstract}




\section{Introduction}

Thermal instability (\citealp{Field1965}) is a fundamental mechanism that governs the formation of a dense gas from a diffuse gas and the evaporation of dense gas in a wide range of astrophysical phenomena.
It arises when radiative cooling locally dominates over heating, leading to runaway condensation or heating depending on the nature of the perturbation.  
A prominent example of thermal instability is the formation of a multi-phase structure in the interstellar medium, where dense, cold neutral medium coexists with diffuse, warm neutral medium (e.g., \citealp{Schwarz_etal1972, Koyama_Inutsuka2000}).
In this context, shock waves compress the warm neutral medium, which then cools and condenses into the cold neutral medium—forming the seeds of molecular clouds  (\citealp{McKee_Ostriker1977, Koyama_Inutsuka2002, Audit_Hennebelle2005, Inutsuka_etal2005, Heitsch_etal2006, Vazquez-Semadeni_etal2007, Hennebelle_Audit2007, Inoue_Inutsuka2008, Inoue_Inutsuka2012}).
Observations of the warm and cold neutral phases of atomic hydrogen support this picture (e.g., \citealp{Dickey_Lockman1990, Heiles_Troland2003, Jenkins_Tripp2011}).
Thermal instability provides a natural explanation for the fragmentation of the interstellar medium into these components.

Thermal instability is also thought to play an important role in the medium surrounding a galaxy (the circum-galactic medium).  
In this environment, cold gas filaments or HI clouds embedded in a hot gas are observed at galactic halo regions (e.g., \citealp{Wakker_etal2012, Tumlinson_etal2017, Putman_etal2012b, McCourt_etal2012, Ledos_etal2024, Ramesh_etal2024, Yao_etal2025}), and thermal instability has been proposed as a viable formation mechanism of the clouds.  
Such clod gas are somtimes named as the intermediate- or high-velocity clouds depending on its proper velocities (e.g., \citealp{Wolfire_etal1995, Binney_etal2009, Lucchini_etal2024, Ramesh_etal2024}), and are considered observational evidence of the baryon cycle within galaxies (e.g., \citealp{Tumlinson_etal2017, Shimoda_etal2024, Shimoda_Asano2024}).
They are often detected via absorption features in quasar spectra (e.g., \citealp{Sembach+2003, Shull+2009, Ashley+2022, Ashley+2024}), suggesting the existence of cool, dense clumps in pressure balance with their hotter surroundings.

On even larger scales, thermal instability may affect the dynamics of the intracluster medium found in galaxy clusters.  
The hot, X-ray-emitting plasma in these systems is susceptible to radiative cooling, potentially leading to the development of cooling flows (e.g., \citealp{Sharma_etal2010, Quataert2011, McCourt_etal2012, Cappiello_Nulsen2015, Meece_etal2015, Beckmann_etal2022, Beckmann2022}). 
Thermal instability has been proposed to explain the formation of cold gas condensations within these flows, and the consequent fueling of star formation or active galactic nuclei. 
These examples demonstrate that thermal instability operates across a wide range of physical conditions and spatial scales, influencing both structure formation and feedback processes in astrophysical systems.


Specifically, we focus on the dynamics of the circum-galactic medium (CGM), where intermediate- and high-velocity clouds are formed from hot gas. 
Previous theoretical studies performed the analyses of thermal instability in a gravitationally stratified medium by  linear analyses and numerical simulations (e.g., \citealp{Defouw1970, Binney_etal2009, Nipoti2010, McCourt_etal2012}). 
The large-scale structures of the stratified medium such as the gradients of pressure and density modifies the behavior of perturbations by buoyancy forces,  which can either suppress or amplify the growth rate of unstable modes. 
Moreover, it has been shown that the thermal instability can be altered to oscillatory growing, over-stable modes due to the buoyancy force from the monotonically growing , unstable modes seen in a uniform medium.  
This implies that the origins and dynamics of the intermediate- and high-velocity clouds as products of the instability are potentially characterized by the structure of CGM via the buoyancy force (\citealp{Defouw1970, McCourt_etal2012}).
However, these studies have left several open questions. 
For example, most previous investigations have relied on simplifying assumptions, such as the Boussinesq or isobaric approximations, which are only valid in the short-wavelength limit.
The growth rates and maximum growth wavelength of the over-stable modes have not been also established yet. 
These are necessary pieces to derive the physical picture of the growth modes and to study the non-linear stage by using numerical simulations with sufficient validities. 
In addition, the previous studies focused exclusively on the convectively stable case, while the convectively unstable case is also important for the accretion dynamics of the cold gas on to the central galaxy(s). 
Thus, further comprehensive studies of thermal instability in the stratified medium are required.

This paper aims to advance the theoretical understanding of thermal instability in a gravitationally stratified medium, with a particular focus on the CGM. 
We achieve this by carefully analyzing the dispersion relation and its underlying approximations. 
By explicitly incorporating the buoyancy frequency into the analysis, we explore how gravitational stratification modifies the growth of instability and identify the conditions under which the most unstable modes arise. 
Additionally, we provide a qualitative discussion on the conditions for over-stability and how the gravitational field affects thermal instability.

The structure of this paper is as follows: 
In Section \ref{sec:1D}, we revisit the dispersion relation for a uniform medium as a reference case for our subsequent analysis. 
We also propose a simpler formula to estimate the most unstable conditions. 
Then, in Section~\ref{sec:2D}, we extend our analysis to a gravitationally stratified medium, where the interplay between thermal instability and convective stability becomes important. 
In Section~\ref{sec:discuss}, we examine the combined effects of thermal and buoyant forces. 
We then use a simplified CGM model with a linear temperature gradient to estimate the typical length and timescales associated with the formation of high-velocity clouds.
This provides an estimation for the minimum resolution required for CGM simulations.
Finally, we summarize our findings and discuss their astrophysical implications in Section~\ref{sec:summary}.

\section{Useful approximation for thermal instability in uniform medium}\label{sec:1D}

This section reviews thermal instability in a spatially uniform medium \citep{Field1965}.
First, in preparation for the case of a gravitationally stratified medium (Section \ref{sec:2D}), we derive the exact dispersion relation by following \citep{Field1965} in Sections \ref{subsec:1D_DR} and \ref{subsec:1D_crit}.
Then, in Sections \ref{subsec:1D_short} and \ref{subsec:1D_long}, we propose more convenient approximations for both shorter and longer wavelength limit.
We also present a method to estimate the most unstable wavelength and its growth rate in Section \ref{subsec:1D_MostUnstable}.
These simple descriptions are useful for quantitatively discussing the motion of fluid elements in a gravitationally-stratified medium in Section \ref{sec:discuss}.

We consider the dynamics of an ideal gas under the influences of radiative cooling and thermal conduction.
The system is described by the following hydrodynamic equations, representing conservation of mass, momentum, and energy, along with the equation of state:
\begin{align}
 	&
	\dfrac{\partial n}{\partial t} + \nabla \cdot [n \bm{v} ] = 0,
	\label{eq:1D:Basic_EoC}\\[3mm]
	&
	\overline{m} n \left [ \dfrac{\partial \bm{v}}{\partial t} + (\bm{v} \cdot \nabla ) \bm{v} \right ] = -\nabla P,
	\label{eq:1D:Basic_EoM}\\[3mm]
	&
	\dfrac{1}{\gamma - 1} \left [ \dfrac{\partial P}{\partial t} + (\bm{v} \cdot \nabla )P \right ] - \dfrac{\gamma P}{(\gamma - 1)n} \left [ \dfrac{\partial n}{\partial t} + (\bm{v} \cdot \nabla )n \right ]
	= -\mathcal{L} (n,T) + \nabla \cdot [\kappa \nabla T],
	\label{eq:1D:Basic_Energy}\\[3mm]
	&
	P = n k_{\rm B} T,
	\label{eq:1D:Basic_EoS}
\end{align}
where $n,\ T$, and $P$ denote the number density, temperature, and pressure, respectively, while $\bm{v}$ represent the velocity field.
The constants $\overline{m},\ \gamma$, and $k_{\rm B}$ represent the mean particle mass, adiabatic index, and Boltzmann constant, respectively. 
The function $\mathcal{L}(n, T)$ represents the net radiative cooling rate per unit volume, and $\kappa$ is the thermal diffusivity.

\subsection{Exact dispersion relation}
\label{subsec:1D_DR}

In this section, we derive the dispersion relation for thermal instability.
We assume a uniform unperturbed state with constant values $n_0$, $T_0$, $P_0$, and $\bm{v}_0 = \bm{0}$ in radiative equilibrium, satisfying $\mathcal{L}(n_0, T_0) = 0$.
The unperturbed state variables are indicated by a subscript ``0''.
Then, we apply the sinusoidal function as the fluctuation, $f_1 = \delta f \exp (\sigma t + i\bm{k} \cdot \bm{x})$, for an unperturbed quantity, $f_0$.
Here, $\sigma$ and $\bm{k}$ denote the growth rate of the instability and the wave number.
Substituting $f = f_0 + f_1$ into Equations \eqref{eq:1D:Basic_EoC} - \eqref{eq:1D:Basic_EoS} and linearizing in the perturbation amplitude, we obtain the following linear equations:
\begin{align}
	&
	\sigma \delta n + i n_0 \bm{k} \cdot \delta \bm{v} = 0,
	\label{eq:1D:Linear_EoC}\\[3mm]
	&
	\sigma \overline{m} n_0 \delta \bm{v} = -i\bm{k} \delta P, 
	\label{eq:1D:Linear_EoM}\\[3mm]
	&
	\dfrac{\sigma}{\gamma - 1} \delta P - \sigma \dfrac{\gamma P_0}{(\gamma - 1)n_0} \delta n = -\mathcal{L}_n \delta n - \mathcal{L}_T \delta T -\kappa k^2 \delta T,
	\label{eq:1D:Linear_Energy}\\[3mm]
	&
	\dfrac{\delta P}{P_0} = \dfrac{\delta n}{n_0} + \dfrac{\delta T}{T_0},
	\label{eq:1D:Linear_EoS}
\end{align}
where $\mathcal{L}_n \equiv (\partial \mathcal{L} / \partial n)_T$ and $\mathcal{L}_T \equiv (\partial \mathcal{L} / \partial T)_n$.
In thermally stable case, where $\partial P / \partial n > 0$, a region with higher density ($\delta n > 0$) has higher pressure ($\delta P > 0$) than its surroundings.
This denser region pushes back on the surrounding medium, returning to its original uniform state.
Conversely, in thermally unstable case, where $\partial P / \partial n < 0$, the internal energy in the denser region decreases more rapidly.
This is because the collision frequency of gas particles becomes larger in the denser region, a process described by the radiative cooling rate $\mathcal{L} (n, T) \propto n^2$.
Consequently, the denser region is compressed further by its surroundings, leading to a runaway increase in density.

To quantitatively discuss in detail, combining Equations \eqref{eq:1D:Linear_EoC} - \eqref{eq:1D:Linear_EoS} yields the dispersion relation as follows:
\begin{align}
	\sigma^3  
	+ \dfrac{\gamma - 1}{n_0 k_{\rm B}}(\mathcal{L}_T + \kappa k^2)\sigma^2 
	+ C_{\rm s}^2k^2 \sigma 
        - \dfrac{(\gamma - 1)C_{\rm s}^2}{\gamma n_0k_{\rm B}}k^2 \left ( \dfrac{n_0}{T_0}\mathcal{L}_n - \mathcal{L}_T - \kappa k^2 \right )
        = 0,
        \label{eq:1D:Linear_DR}
\end{align}
where $C_{\rm s} = \sqrt{\gamma P_0 / \overline{m} n_0}$ denotes the sound speed.

To simplify this expression, we define three characteristic wave numbers following \citet{Field1965}:
\begin{align}
	k_n \equiv \dfrac{(\gamma - 1)\mathcal{L}_n}{k_{\rm B}T_0C_{\rm s}},\quad
    	k_T \equiv \dfrac{(\gamma - 1)\mathcal{L}_T}{n_0k_{\rm B}C_{\rm s}},\quad
    	k_K \equiv \dfrac{n_0k_{\rm B}C_{\rm s}}{(\gamma - 1)\kappa}.
	\label{eq:1D:Linear_k}
\end{align}
$k_n,\ k_T$, and $k_K$correspond to the wave numbers associated with isothermal and isochoric modes, and the inverse of the conductive mean free path.
Substituting them into Equation~\eqref{eq:1D:Linear_DR} yields the dimensionless form:
\begin{align}
	\sigma^3 + C_{\rm s} \left ( k_T + \dfrac{k^2}{k_K} \right ) \sigma^2 + C_{\rm s}^2 k^2 \sigma 
	- \dfrac{C_{\rm s}^3 k^2}{\gamma} \left ( k_n - k_T - \dfrac{k^2}{k_K} \right ) = 0.
	\label{eq:1D:Linear_DR:k}
\end{align}

We further normalize this equation using $k_n$ and $C_{\mathrm{s}}$:
\begin{align}
	&
	\tilde{\sigma}^3 + (\alpha + \beta \tilde{k}^2) \tilde{\sigma}^2 + \tilde{k}^2 \tilde{\sigma} - \gamma^{-1}\tilde{k}^2 (1 - \alpha - \beta \tilde{k}^2) = 0.
	\label{eq:1D:Linear_DR:norm}\\[1mm]
	&
	\tilde{\sigma} \equiv \dfrac{\sigma}{k_n C_{\rm s}} , \quad 
	\tilde{k} \equiv \dfrac{k}{k_n}, \quad
	\alpha \equiv \dfrac{k_T}{k_n} = \dfrac{T_0 \mathcal{L}_T}{n_0 \mathcal{L}_n}, \quad
	\beta \equiv \dfrac{k_n}{k_K} = \dfrac{(\gamma - 1)^2 \kappa \mathcal{L}_n}{n_0 k_\mathrm{B}^2 T_0 C_{\rm s}^2},
	\label{eq:1D:Linear_DR:norm_values}
\end{align} 
the variables with the tilde notation (e.g.., $\tilde{\sigma}$ or $\tilde{k}$) denote normalized quantities. 
Thus, the dispersion relation depends on the three dimensionless parameters $\alpha$, $\beta$, and $\gamma$.

In the interstellar medium, the net radiative cooling rate is often expressed as $\mathcal{L}(n, T) = n^2 \mathit\Lambda(T) - n \mathit\Gamma$, where $\mathit\Lambda(T)$ denotes the cooling function and $\mathit\Gamma$ is the radiative heating rate.  
Under this formulation, the parameter $\alpha$ can be rewritten as $\alpha = d \ln \mathit\Lambda(T) / d \ln T$.  
If the cooling function follows a power-law form, i.e., $\mathit\Lambda(T) \propto T^\alpha$, then $\alpha$ corresponds directly to the power-law index.
Equation~\eqref{eq:1D:Linear_DR:norm} is a cubic equation in the dimensionless growth rate $\tilde{\sigma}$, and the system becomes thermally unstable when the constant term is negative.  
This condition is written as $- \gamma^{-1} \tilde{k}^2 (1 - \alpha - \beta \tilde{k}^2) < 0$.  
Neglecting the effect of thermal conduction ($\beta = 0$), the instability criterion simplifies to $\alpha < 1$, which is equivalent to the well-known Field criterion, i.e., $(\partial \mathcal{L} / \partial T)_P < 0$ (\citealp{Field1965}), and $\partial P / \partial n < 0$.

\begin{figure}
	\centering
	\includegraphics[width=\linewidth]{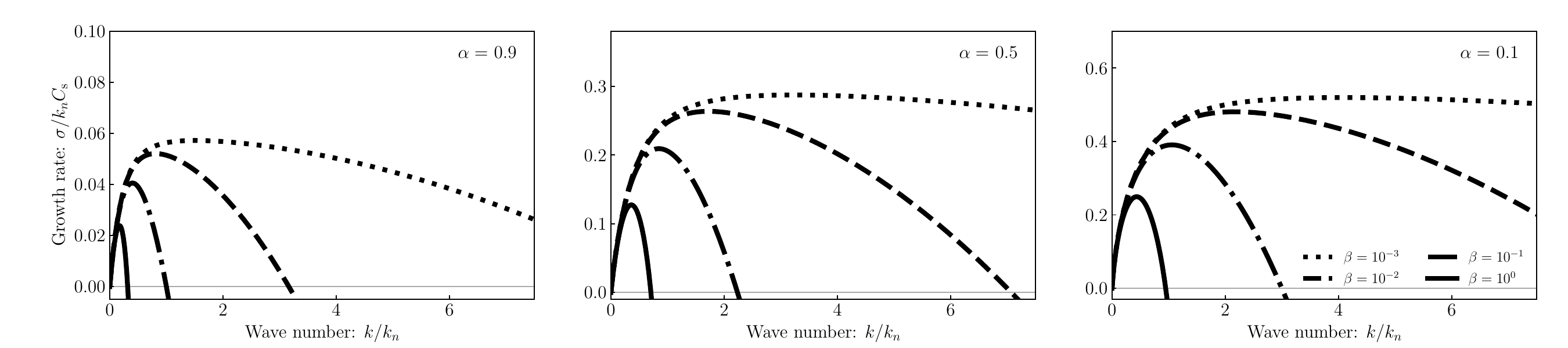}
	\caption{The growth rates as a function of the wave number for $\alpha = 0.9,\ 0.5$, and 0.1.
	The results are shown for a fixed $\gamma = 5/3$ with varying the effect of thermal conduction, $\beta = 10^{-3},\ 10^{-2},\ 10^{-1},\ 10^0$.
	The horizontal and vertical axes are the normalized wave number and growth rate of the unstable mode.
	The difference of lines denotes the difference of the effects of the thermal conduction, $\beta$.}
	\label{fig:1D:Linear_DR}
\end{figure}

Figure~\ref{fig:1D:Linear_DR} shows the growth rate as a function of the wave number $k$, obtained by solving Equation~\eqref{eq:1D:Linear_DR:norm} for $\alpha = 0.9, 0.5$, and 0.1, with $\gamma = 5/3$.  
The effect of thermal conduction is varied through the parameter $\beta$, taking values of $10^{-3}$, $10^{-2}$, $10^{-1}$, and $10^0$.  
In this case, all solutions are purely unstable modes with real-valued growth rates, i.e., $\sigma \in \mathbb{R}$.

\subsection{Critical wavelength}
\label{subsec:1D_crit}

Thermal instability in a uniform medium has a critical wavelength below which it is stabilized due to the conduction. In the dispersion relation, Equation \eqref{eq:1D:Linear_DR:norm}, this effect is expressed as the sign of the last term: the negative (positive) sign results in the growth (decay) of perturbation. 
Thus, the critical length is defined by 
\begin{align}
	1 - \alpha - \beta \tilde{k}_{\mathrm{crit}}^2 = 0.
\end{align}
The corresponding critical wavelength is expressed as
\begin{align}
	\lambda_{\mathrm{F}} \equiv 2\pi \sqrt{\kappa \left| \frac{\partial \mathcal{L}}{\partial T}\right|_P^{-1}  },
	\label{fig:1D:Linear_FieldLength}
\end{align}
where $\lambda_{\mathrm{F}} \equiv 2\pi / k_{\mathrm{crit}}$ is known as the Field length \citep{Field1965}.
The perturbation with a shorter wavelength than $\lambda_{\rm F}$ is diffused out by the thermal conduction before changing the net radiative rate $(\partial \mathcal{L}/\partial T)_P$.

\subsection{Short wavelength approximation}
\label{subsec:1D_short}

The full dispersion relation derived in Section~\ref{subsec:1D_DR} is a cubic equation and somewhat complex to solve.  
To obtain a simpler form that is easier to interpret, we consider an approximation valid in the short-wavelength regime.
We consider a short wavelength perturbation that the sound crossing time for the given wavelength is much shorter than the cooling time.
Since the pressure perturbation is only given by the sound wave gone away from the perturbed region , the isobaric condition of $\delta P \approx 0$ is approximately achieved there. 
In this approximation,  the perturbed region is dominated by the density and temperature fluctuations with an opposite phase ($\delta P/P = \delta n/n + \delta T/T \approx 0)$.
We can derive a reduced dispersion relation from Equations \eqref{eq:1D:Linear_Energy} - \eqref{eq:1D:Linear_EoS} under the condition of $\delta P = 0$ as Equation \eqref{eq:1D:Linear_DR:short}, where the first two-terms of Equation \eqref{eq:1D:Linear_DR:norm} are vanished by the approximation. Note that the equations \eqref{eq:1D:Linear_EoC} - \eqref{eq:1D:Linear_EoM} are not used because we neglect the sound wave:
\begin{align}
	\tilde{\sigma} = \gamma^{-1} (1 - \alpha - \beta \tilde{k}^2).
	\label{eq:1D:Linear_DR:short}
\end{align}

\begin{figure}
	\centering
	\includegraphics[width=\linewidth]{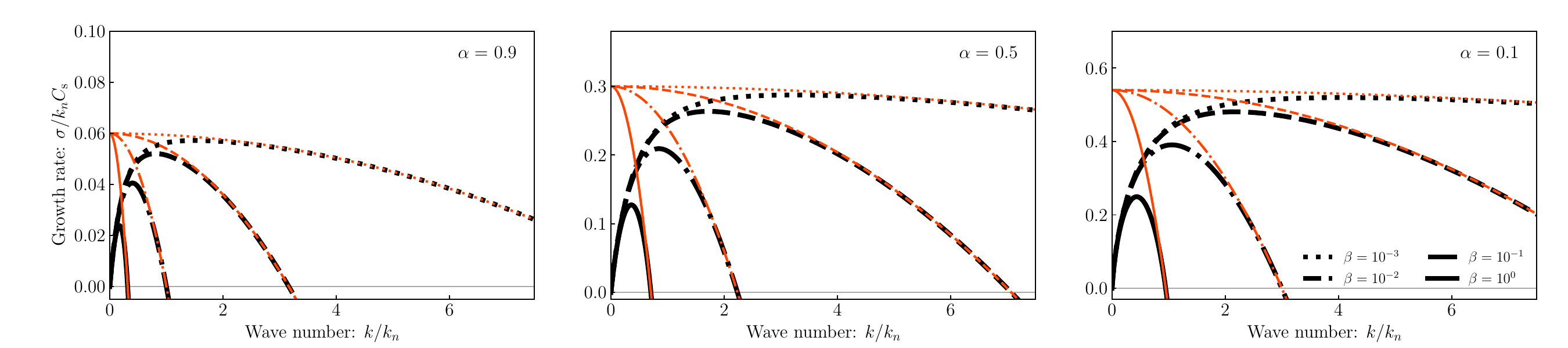}
	\caption{The dispersion relation under the short wavelength approximation in Equation \eqref{eq:1D:Linear_DR:short}. 
	The orange lines represent the approximate solutions, while the black solid lines are the exact solutions as given by Equation \eqref{eq:1D:Linear_DR}.}
	\label{fig:1D:Linear_DR:short}
\end{figure}

Figure~\ref{fig:1D:Linear_DR:short} shows the growth rate as a function of wave number under this approximation, compared to the exact solution.  
Our approximation accurately reproduces the behavior of the full solution near the most unstable wave number and on the short-wavelength side.
The deviation from the exact solution begins around the most unstable wavelength scale.  To catch the most unstable wavelength, we should consider the other approximation valid in the long-wavelength side (Section \ref{subsec:1D_long}).
Equation~\eqref{eq:1D:Linear_DR:short} yields a finite value $\tilde{\sigma} = \gamma^{-1}(1 - \alpha)$, corresponding to the characteristic cooling rate of the system, which is independent of wave number.

\subsection{Long wavelength approximation}
\label{subsec:1D_long}

In this section, we consider the opposite limit: the long-wavelength regime.
This regime is defined by the condition that the sound-crossing time across the perturbed wavelength, $t_{\rm sound}$, is significantly longer than the typical cooling time, $t_{\rm cool}$ ($t_{\rm sound} \gg t_{\rm cool}$).
In this case, any deviations from the unperturbed state due to internal energy (i.e., entropy) fluctuations are rapidly adjusted to the equilibrium state, while sound waves (pressure fluctuations) are allowed to develop.
This situation corresponds to perturbations that also follow the thermal equilibrium condition $\delta \mathcal{L} (n, T) = 0$.
Therefore, the perturbed net cooling and thermal conduction satisfy the condition $\mathcal{L}_n \delta n + \mathcal{L}_T \delta T + \kappa k^2 \delta T= 0$, which approximately corresponds to situations without the internal energy fluctuations (entropy fluctuations).  
Although the condition $\mathcal{L}_n \delta n + \mathcal{L}_T \delta T = 0$ corresponds to the long wavelength approximation, we include the thermal conduction term in this regime to examine its specific influence on the dispersion relation.
The reduced form of the growth rate can be derived from Equations \eqref{eq:1D:Linear_EoC} - \eqref{eq:1D:Linear_EoS} under this condition as
\begin{align}
	\tilde{\sigma}^2 = \gamma^{-1} \dfrac{1 - \alpha  - \beta \tilde{k}^2}{\alpha + \beta \tilde{k}^2}\tilde{k}^2,
	\label{eq:1D:Linear_DR:long}
\end{align}
where the positive root of $\tilde{\sigma}$ represents the unstable solution.

\begin{figure}
	\centering
	\includegraphics[width=\linewidth]{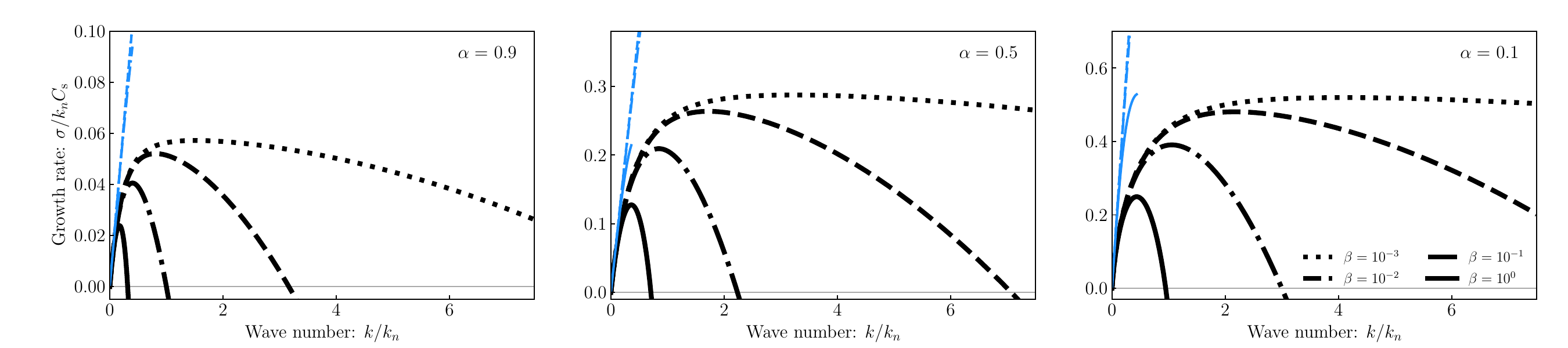}
	\caption{The dispersion relation under the long wavelength approximation, i.e., $\mathcal{L}_n \delta n + \mathcal{L}_T \delta T + \kappa k^2 \delta T = 0$ in Equation \eqref{eq:1D:Linear_DR:long}.
	The sky blue lines represent the approximate solutions, while the black solid lines are the exact solutions.}
	\label{fig:1D:Linear_DR:long}
\end{figure}

Figure \ref{fig:1D:Linear_DR:long} compares this approximation with the exact dispersion relation. 
The approximation is valid and accurate in the long-wavelength regime. 
In the extreme limit of long-wavelength perturbations, the thermal conduction term also becomes negligible, meaning the growth rates do not depend on $\beta$.
This allows us to further simplify the dispersion relation to Equation \eqref{eq:1D:Linear_DR:long2}:
\begin{align}
	\tilde{\sigma}^2 = \gamma^{-1} \dfrac{1 - \alpha}{\alpha}\tilde{k}^2.
	\label{eq:1D:Linear_DR:long2}
\end{align}
Hereafter, we refer to Equation \eqref{eq:1D:Linear_DR:long2} as the long-wavelength approximation.
Note that this approximation is not equivalent to the adiabatic solution, but rather to the isothermal one. 
This implies that cooling is much more dominant than the adiabatic change. 
Thus, when we assume the polytropic relation $\delta P / P_0 = \gamma_{\rm eff}\cdot \delta n / n_0$, i.e., $P \propto n^{\gamma_{\rm eff}}$, the effective adiabatic index ($\gamma_{\rm eff}$) is defined as $\gamma_{\rm eff} = (\alpha - 1)/\alpha$ ( $\neq \gamma = 5/3$ for monoatomic gas) under this approximation [Equation \eqref{eq:1D:Linear_EoS} and $\mathcal{L}_n \delta n + \mathcal{L}_T \delta T = 0$ are used].

\subsection{Estimation of the most unstable condition}
\label{subsec:1D_MostUnstable}

Combining the approximations in the short- (Section \ref{subsec:1D_short}) and long-wavelength (Section \ref{subsec:1D_long}) perturbations, we now estimate the most unstable wavelength and growth rate, which lies between the short- and long-wavelength regimes.  
Heuristically, we derive the maximum growth rate by simply combining the dispersion relations represented by \eqref{eq:1D:Linear_DR:short} and \eqref{eq:1D:Linear_DR:long2}, and estimating the peak of its growth rate.
The dispersion relation that combines the short- and long-wavelength approximations is as follows:
\begin{align}
	\alpha \tilde{\sigma}^2 + \tilde{k}^2 \tilde{\sigma} - \gamma^{-1} \tilde{k}^2 \left( 1 - \alpha - \beta \tilde{k}^2 \right) = 0.
	\label{eq:1D:Linear_DR:most}
\end{align}
To identify the most unstable wave number $k_{\mathrm{max, TI}}$, we differentiate the equation with respect to $k^2$ and impose the condition $d\sigma/dk = 0$, which leads to
\begin{align}
	\tilde{\sigma} = \gamma^{-1} \left( 1 - \alpha - 2\beta \tilde{k}_{\rm max, TI}^2 \right).
	\label{eq:1D:Linear_DR:most_derivative1}
\end{align}
Substituting this into Equation~\eqref{eq:1D:Linear_DR:most} yields
\begin{align}
	\alpha \tilde{\sigma}^2 = \gamma \beta \tilde{k}_{\rm max, TI}^4.
	\label{eq:1D:Linear_DR:most_derivative2}
\end{align}
which can be solved to obtain
\begin{align}
	&
	\tilde{k}_{\rm max, TI}^2 \approx (1 - \alpha) \left[ 2\beta + \sqrt{ \dfrac{\gamma \beta}{\alpha} } \right]^{-1} 
	= \tilde{k}_{\rm crit}^2 \left[ 2 + \sqrt{ \dfrac{\gamma}{ \alpha \beta} } \right]^{-1},
	\label{eq:1D:Linear_DR:most_wn} \\[1mm]
	&
	\lambda_{\rm max, TI} \equiv \dfrac{2\pi}{k_{\rm max, TI}} 
	\approx \lambda_{\rm F} \left[ 2 + \dfrac{n_0 k_{\rm B} C_{\rm s}}{\gamma - 1} \sqrt{ \dfrac{\gamma}{\kappa \mathcal{L}_T } } \right]^{1/2}.
	\label{eq:1D:Linear_DR:most_wl}
\end{align}
By substituting $\tilde{k}_{\mathrm{max, TI}}$ back into Equation~\eqref{eq:1D:Linear_DR:most_derivative2}, we also estimate the most unstable growth rate, $\sigma_{\rm max, TI}$.

\begin{figure}
	\centering
	\includegraphics[width=\linewidth]{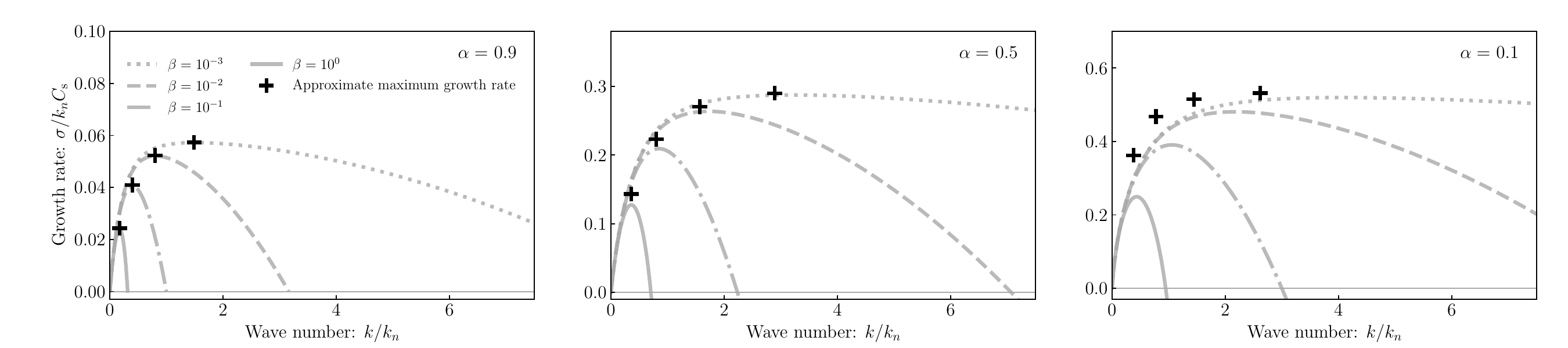}
	\caption{Approximation for the maximum growth rate and the most unstable wavelength $(k_{\rm max, TI},\ \sigma_{\rm max, TI})$ for each parameters. The grey curves show the exact dispersion relations, and the cross markers indicate the approximate estimates obtained from Equations \eqref{eq:1D:Linear_DR:most_derivative1} and \eqref{eq:1D:Linear_DR:most_wn}.}
	\label{fig:1D:Linear_DR:max}
\end{figure}

Figure~\ref{fig:1D:Linear_DR:max} compares the approximate result with the exact dispersion relation.  
The grey curves show the exact solution, while the cross symbols represent the estimated most unstable wave number and growth rate for each parameters.  
The estimated values of the most unstable condition successfully reproduce the peak of the exact dispersion relation, especially for $\alpha \sim 1$.
Although it deviates from the peak of the exact solution when $\alpha < 1$, the most unstable wavelength remains consistent.
This overestimation of the growth rate is likely due to the partial neglect of the thermal conduction term.

\section{Application in the gravitationally stratified medium}
\label{sec:2D}
In this section, we derive the dispersion relation and its approximate fomula for thermal instability in the gravitationally-stratified medium.
We adopt a two-dimensional Cartesian coordinate system $(x, z)$, where $z$ denotes the vertical direction.
From Sections \ref{subsec:2D:Unp} to \ref{subsec:2D:DR}, we derive the exact dispersion relation by linearizing the basic hydrodynamic equations around the unperturbed state.
We analyze the convectively unstable case in Section \ref{subsec:discuss:CIy}, and then the stable case in Section \ref{subsec:discuss:CIn}.
In these sections, we fix the thermal instability parameters ($\alpha = 0.5$, $\beta = 0.01$, and $\gamma = 5/3$) to focus on the effects of stratification.
Finally, as in Section \ref{sec:1D}, we adapt the approximations for both short- and long-wavelength limits in Section \ref{subsec:2D:Approx} and estimate the most unstable condition in Section \ref{subsec:discuss:MUC}.

We consider the dynamics of an ideal gas under the influences of gravitational acceleration, radiative cooling and thermal conduction. 
The governing hydrodynamic equations are given by:
\begin{align}
	&
	\dfrac{\partial n}{\partial t} + \nabla \cdot  [n \bm{v}] = 0,
	\label{eq:2D:Basic_EoC}\\[3mm]
	&
	\overline{m} n \left [ \dfrac{\partial \bm{v}}{\partial t} + (\bm{v} \cdot \nabla) \bm{v}  \right ] = -\nabla P - \overline{m} n g \bm{e}_z,
	\label{eq:2D:Basic_EoM}\\[3mm]
	&
	\dfrac{1}{\gamma - 1} \left [ \dfrac{\partial P}{\partial t} + (\bm{v} \cdot \nabla )P \right ] - \dfrac{\gamma P}{(\gamma - 1)n} \left [ \dfrac{\partial n}{\partial t} + (\bm{v} \cdot \nabla )n \right ]
	= -\mathcal{L} (n,T) + \nabla \cdot [\kappa \nabla T] - \overline{m} n g\bm{e}_z \cdot \bm{v},
	\label{eq:2D:Basic_Energy}
\end{align}
Here, $\bm{e}_z$ is the unit vector in the vertical ($z$) direction, and $g > 0$ denotes the constant gravitational acceleration pointing downward (i.e., in the $-z$-direction).
We use Equations~\eqref{eq:2D:Basic_EoC}--\eqref{eq:2D:Basic_Energy} as the foundation for the linear stability analysis presented in this section.

\subsection{Unperturbed state}
\label{subsec:2D:Unp}

We define the unperturbed state in preparation for the linear stability analysis presented in Section~\ref{subsec:2D:Linear}.  
We assume that the background medium is uniform in the horizontal $x$-direction, and in hydrostatic and energy equilibrium in the vertical $z$-direction.  
Under these assumptions, the unperturbed quantities satisfy the following one-dimensional ordinary differential equations (ODEs), along with the equation of state given in Equation~\eqref{eq:1D:Basic_EoS}:
\begin{align}
	&
	 -\dfrac{dP_0}{dz} -\overline{m}n_0 g = 0,
	\label{eq:2D:Unp_EoM}\\[3mm]
	&
	-\mathcal{L} (n_0, T_0) + \dfrac{d}{dz} \left ( \kappa \dfrac{dT_0}{dz} \right ) = 0.
	\label{eq:2D:Unp_Energy}
\end{align}
In a medium with spatial variations, its structure is characterized by the scale heights.
The inverse scale heights of density $H_n^{-1}$ and pressure $H_P^{-1}$ are defined as
\begin{align}
	&
	\dfrac{1}{H_n} \equiv -\dfrac{1}{n_0} \dfrac{dn_0}{dz} > 0,
	\label{eq:2D:DR_Hn} \\[1mm]
	&
	\dfrac{1}{H_P} \equiv -\dfrac{1}{P_0} \dfrac{dP_0}{dz} > 0,
	\label{eq:2D:DR_H}
\end{align}
where the positivity reflects the fact that both density and pressure decrease with height in hydrostatic equilibrium.
The scale height of temperature $H_T$ is defined similarly, but it can be either positive or negative depending on the temperature gradient.

\subsection{Linearization}
\label{subsec:2D:Linear}

The unperturbed state varies in the vertical ($z$) direction, as described by Equations~\eqref{eq:2D:Unp_EoM} and \eqref{eq:2D:Unp_Energy}.  
In a medium with spatial variations, a precise analysis becomes too complicated. 
The perturbation wavelength is modulated due to the variations of medium as seen in the refraction for example. 
This means that the exact properties of the perturbation are no longer expressed by the simple sinusoidal functions unlike the case of a uniform medium.
However, it is well established that the essential behavior of instabilities can still be captured by assuming sinusoidal perturbations and focusing on the real part of the resulting dispersion relation (e.g., \citealp{Seno_Inutsuka2025conv}).
This approximation becomes particularly appropriate when the scale height of the stratified medium is much larger than the typical length scale of a phenomenon of interest.
In this case, the perturbation behaviors can be expressed approximately even by the simple sinusoidal functions, where the wave numbers varies gradually over space (see, e.g., the linear analyses of convection; \citealp{Chandrasekhar1961hydrodynamic, Mihalas_Mihalas1999, Seno_Inutsuka2025conv}). 
This is known as the WKB approximation.
 We adopt it for the perturbations as $f_1 = \delta f \exp(\sigma t + i k_x x + i k_z z)$, where $k_x$ and $k_z$ are the wave numbers in the horizontal and vertical directions, respectively.  
The two-dimensional analysis decomposing the parallel and perpendicular directions to the stratification is sufficient without loss of generality.
Substituting the perturbed quantities into Equations~\eqref{eq:2D:Basic_EoC}--\eqref{eq:2D:Basic_Energy} and retaining only first-order terms, we obtain the linearized equations:
\begin{align}
	&
	\sigma \delta n + i n_0 k_x \delta v_x + in_0 k_z \delta v_z + \dfrac{dn_0}{dz} \delta v_z = 0,
	\label{eq:2D:Linear_EoC}\\[3mm]
	&
	 \sigma \overline{m} n_0 \delta v_x = - ik_x \delta P,
	 \label{eq:2D:Linear_EoMx}\\[3mm]
	 &
    	\sigma \overline{m} n_0 \delta v_z = - ik_z \delta P  - \overline{m} g \delta n,
   	 \label{eq:2D:Linear_EoMz} \\[3mm]
	 &
    	\frac{1}{\gamma - 1} \left( \sigma \delta P + \frac{dP_0}{dz} \delta v_z \right)  -  \frac{\gamma P_0}{(\gamma - 1)n_0} \left( \sigma \delta n + \frac{dn_0}{dz} \delta v_z \right)
      	=  
    	- \mathcal{L}_n \delta n - \mathcal{L}_T \delta T - \kappa k^2 \delta T - \overline{m} n_0 g \delta v_z,
	\label{eq:2D:Linear_Energy}
\end{align}
Here, we have separated the momentum equations into their horizontal ($x$) and vertical ($z$) components for clarity.

In comparison of the uniform medium case, Equations \eqref{eq:1D:Linear_EoC} -- \eqref{eq:1D:Linear_Energy}, the gravitational acceleration and the gradients of unperturbed sate appear.
The former drives the convection and modifies the thermal instability. 
The latter results in the anisotropic behavior of perturbations along the stratification direction, $z$. 
Even if we consider a perturbation with $\bm{k}=(k_x,0,0)$, the gravitational acceleration affects the thermal instability via $g \delta v_z \exp (\sigma t+ik_x x)$ in the energy equation \eqref{eq:2D:Linear_Energy}.  
This also implies that the perturbation behavior is determined by the combination of the thermal instability and convection. 
For example, an expansion due to the buoyant force prevents  the fluid from the condensation due to the thermal instability.

\subsection{Dispersion relation}
\label{subsec:2D:DR}

\begin{table}
	\centering
	\caption{The definitions of variables related to Section \ref{subsec:2D:DR}.}  
	\label{tab:2D:DR_variables} 
	\begin{tabular}{llc} 
	\hline 
	Symbol 				& 	Definition 																	&	Equation\\
	\hline \hline
	$\sigma$ 				&	Growth rate of instability  														&						\\[1mm]
	$k_x$				&	Wave number in the $x$--direction  												&						\\[1mm]
	$k_z$				&	Wave number in the $z$--direction  												&						\\[1mm]
	$k^2$				&	Squared wave number defined as $k^2 = k_x^2 + k_z^2$ 								&						\\[1mm]
	$\gamma$			&	The adiabatic index  															&						\\[1mm]	
	$C_{\rm s}$			&	Sound speed																&						\\[1mm]
	$g$					&	Absolute value of gravitational acceleration														&						\\[1mm]
	$k_n$				&	The wave number associated with isothermal wave  									&	\eqref{eq:1D:Linear_k}	\\[1mm]
	$k_T$				&	The wave number associated with isochoric wave 									&	\eqref{eq:1D:Linear_k}	\\[1mm]
	$k_K$				&	The inverse of the conductive mean free path 										&	\eqref{eq:1D:Linear_k}	\\[1mm]
	$\alpha$				& 	Dimensionless parameter related to the energy loss rate $\mathcal{L}$, i.e., $k_T / k_n$ 		&	\eqref{eq:1D:Linear_DR:norm_values}\\[1mm]
	$\beta$				&	Dimensionless parameter related to the thermal conduction, i.e., $k_n / k_K$ 				&	\eqref{eq:1D:Linear_DR:norm_values}\\[1mm]
	$\tilde{f}$				&	Physical value $f$ normalized by $C_{\rm s}$ and $k_n$				 				&	\eqref{eq:1D:Linear_DR:norm_values}\\[1mm]
	$H_n$				&	Density scale-height 															&	\eqref{eq:2D:DR_Hn}	\\[1mm]
	$H_P$				&	Pressure scale-height  														&	\eqref{eq:2D:DR_H}	\\[1mm]
	$\omega_{\rm BV}^2$	&	Squared $Brunt-V\ddot{a}is\ddot{a}l\ddot{a}$ frequency 								&	\eqref{eq:2D:DR_BV} 	\\[1mm]
	\hline
	\end{tabular}
\end{table}

In this section, we derive the dispersion relation in this hydrostatic system.
The definitions of the variables related to this section are as shown in Table~\ref{tab:2D:DR_variables}.
Using the linearized equations~\eqref{eq:1D:Linear_EoS} and \eqref{eq:2D:Linear_EoC}--\eqref{eq:2D:Linear_Energy}, we obtain the real part of the dispersion relation as
\begin{align}
	\tilde{\sigma}^4 + (\alpha + \beta \tilde{k}^2 ) \tilde{\sigma}^3 + \left[ \tilde{k}^2 + \frac{1}{\gamma \tilde{H}_n \tilde{H}_P}  \right] \tilde{\sigma}^2 
	- \left[ \gamma^{-1} \tilde{k}^2 (1 - \alpha - \beta \tilde{k}^2) - \frac{1}{\gamma \tilde{H}_n \tilde{H}_P} (\alpha + \beta \tilde{k}^2 ) \right] \tilde{\sigma}  + \tilde{\omega}_{\rm BV}^2 \tilde{k}_x^2 = 0,
	\label{eq:2D:DR}
\end{align}
where $k^2 \equiv k_x^2 + k_z^2$, and all quantities are normalized in the same manner as in Section~\ref{subsec:1D_DR}.
The quantity $\omega_{\mathrm{BV}}^2$ is the square of the $Brunt-V\ddot{a}is\ddot{a}l\ddot{a}$ frequency, which characterizes convective stability:
\begin{align}
	\omega_{\rm BV}^2 = \frac{g}{ H_n} \left( 1 - \frac{2-\gamma}{\gamma} \frac{H_n}{H_P} \right).
	\label{eq:2D:DR_BV}
\end{align}
When $\mathcal{L} = 0$, $\omega_{\mathrm{BV}}^2 > 0$ presents that it is convectively stable and supports oscillatory buoyant motion, while $\omega_{\mathrm{BV}}^2 < 0$ indicates convective instability.
We assume that the density and pressure scale heights are much larger than the typical length scale $\sim k_n^{-1}$.  
This allows us to treat the inverse scale heights as small parameters: $\tilde{H}_n^{-1} \equiv (k_n H_n)^{-1}$ and $\tilde{H}_P^{-1} \equiv (k_n H_P)^{-1}$.  
Neglecting second-order terms associated with $(H_n H_P)^{-1}$, the dispersion relation simplifies to
\begin{align}
	& 
	\tilde{\sigma}^4 + (\alpha + \beta \tilde{k}^2 ) \tilde{\sigma}^3 + \tilde{k}^2 \tilde{\sigma}^2 
	- \gamma^{-1} \tilde{k}^2 (1 - \alpha - \beta \tilde{k}^2) \tilde{\sigma}  + \tilde{\omega}_{\rm BV}^2 \tilde{k}_x^2 = 0,
	\label{eq:2D:DR_WKB} \\[1mm]
	\Longrightarrow\ & 
	\tilde{\sigma} [ \tilde{\sigma}^3 + (\alpha + \beta \tilde{k}^2) \tilde{\sigma}^2 + \tilde{k}^2 \tilde{\sigma} - \gamma^{-1}\tilde{k}^2 (1 - \alpha - \beta \tilde{k}^2) ]
	+ \tilde{\omega}_{\rm BV}^2 \tilde{k}_x^2 = 0.
	\label{eq:2D:DR_WKB2}
\end{align}
The term in brackets in Equation~\eqref{eq:2D:DR_WKB2} is identical to the dispersion relation for thermal instability in a uniform medium, given in Equation~\eqref{eq:1D:Linear_DR:norm}.

\subsection{Convectively unstable case ($\omega_{\rm BV}^2 < 0$)}
\label{subsec:discuss:CIy}

\begin{figure}
	\centering
	\includegraphics[width=0.5\linewidth]{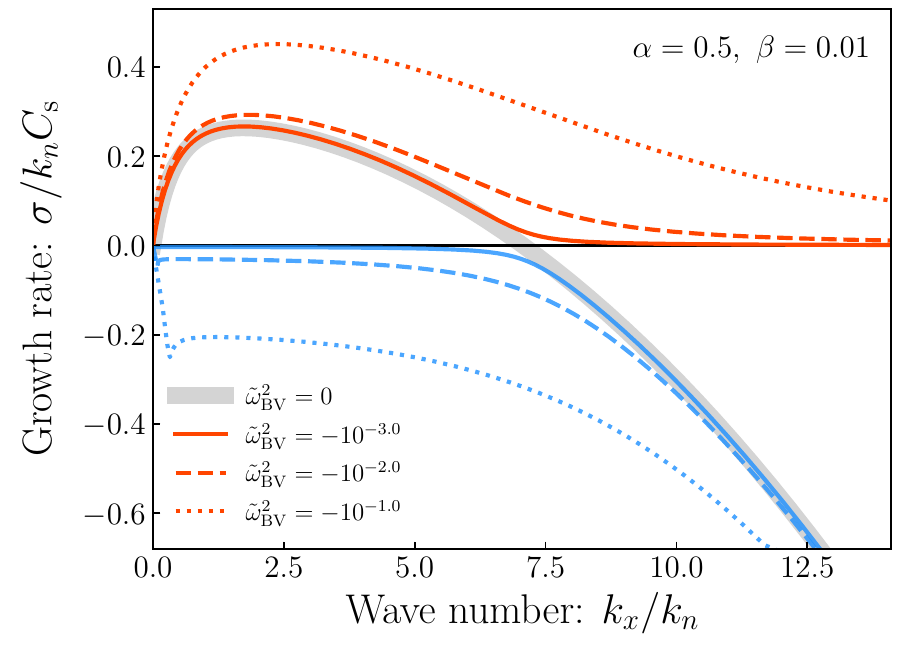}
	\caption{Dispersion relation in the convectively unstable case, for parameters $\alpha = 0.5$, $\beta = 0.01$, $\gamma = 5/3$, and $k_z = 0$.  
	The grey solid line shows the result for a uniform medium, while the colored lines represent the behavior in a gravitationally stratified medium.  
	The red curves correspond to thermally unstable modes in the uniform case, while the blue curves represent modes that originate from vertical oscillations governed by Equation~\eqref{eq:2D:Linear_EoMz}.  
	Different line styles reflect different values of the normalized $Brunt-V\ddot{a}is\ddot{a}l\ddot{a}$ frequency $\tilde{\omega}_{\mathrm{BV}}^2$.
	Note that larger wavenumber modes are stabilized by viscosity as shown in Section \ref{subsubsec:discuss:viscosity}.}
	\label{fig:discuss_DR:Cy}
\end{figure}

In this section, we consider the convectively unstable case ($\omega_{\rm BV}^2 < 0$).
Figure~\ref{fig:discuss_DR:Cy} presents the dispersion relation obtained by solving Equation~\eqref{eq:2D:DR_WKB2}.  
It can be seen that thermal instability is enhanced with increasing the gravitational acceleration ($\omega^2_{\rm BV}\propto g$).
In particular, the unstable growth of thermal instability is seen even for perturbations with $\lambda < \lambda_{\rm F}$ (the Field length) due to the gravity, at which perturbations in the uniform medium become stable due to the thermal conduction.  
This behavior arises because convective instability cannot be stabilized by thermal conduction alone (c.f., \citealp{Seno_Inutsuka2025conv}).  
The stabilization of the convective instability appears due to the viscosity and thermal conduction as studied by \cite{Seno_Inutsuka2025conv}. 
In section \ref{subsubsec:discuss:viscosity}, we show how introduce the effects of viscosity and more realistic growth rate.
There also are stable modes with negative growth rates ($\sigma<0$, blue curves). 
The negative growth rate indicates  a quick decay of perturbation and the decay rates increase under strong gravity with $\omega_{\rm BV}^2<0$. 
Thus, we find that only one unstable mode exists in this convectively unstable case. 
Through following analysis of the dispersion relation, we further find that this unstable mode corresponds to the classical thermally unstable mode.

\subsubsection{Critical wavelength with the effect of viscosity}
\label{subsubsec:discuss:viscosity}
As Figure \ref{fig:discuss_DR:Cy} suggests, the thermally unstable mode becomes unstable at all wavelengths, including those shorter than the Field length.
This is because convective instability persists even for perturbations with wavelengths shorter than the Field length.
However, convective instability can be stabilized only by the combination of both viscosity and thermal conduction (e.g., \citealp{Seno_Inutsuka2025conv}).
To determine the critical wavelength in this case, we must include the effect of viscosity in our analysis.
Thus, in this section, we examine the critical wave number $k_{\rm crit}$ of the marginal condition ($\sigma=0$) with considering the effect of viscosity.

\begin{figure}
	\centering
	\includegraphics[width=0.5\linewidth]{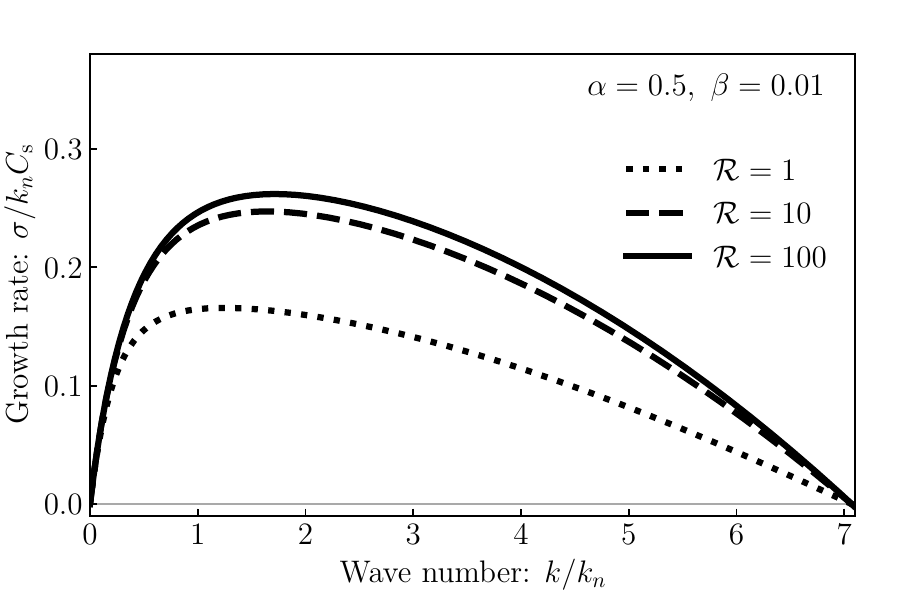}
	\caption{The dispersion relation including viscosity in the uniform medium. 
	The discrepancies among the lines originate from differences in the Reynolds number $\mathcal{R} = 1, 10, 100$.}
	\label{fig:discuss:visc_1D}
\end{figure}

\paragraph*{In uniform medium: }
First, we introduce the effect of viscosity on thermal instability in a uniform medium. 
Since Unperturbed state is stationary, replacing $\sigma$ with $\sigma + \nu k^2$ in Equations \eqref{eq:1D:Linear_EoM}. This substitution is equivalent to modifying the momentum equation to $\sigma \bar{m} n_0 \delta \bm{v} = -i\bm{k} \delta P - \bar{m} n_0 \nu k^2 \delta \bm{v}$ (while other linearized equations retain their original form).
Reanalyzing the system with this modified equation yields the following dispersion relation:
\begin{align}
	\tilde{\sigma}^3 + \{ \alpha + (\beta + \mathcal{R}^{-1}) \tilde{k}^2 \} \tilde{\sigma}^2
	+ \{ 1 + (\alpha + \beta \tilde{k}^2) \mathcal{R}^{-1} \} \tilde{k}^2 \tilde{\sigma} - \gamma^{-1} \tilde{k}^2 (1 - \alpha - \beta \tilde{k}^2) = 0.
	\label{subsec:discuss:uniform_viscosity_DR} 
\end{align}
Here, $\mathcal{R}^{-1} \equiv k_n / k_V$, where $k_V \equiv C_{\rm s} / \nu$.
Thus, $\mathcal{R}$ corresponds to the Reynolds number when adapting $k_n^{-1}$ and $C_{\rm s}$ to the typical length and speed.
The important point is that the last term $\gamma^{-1} \tilde{k}^2(1-\alpha -\beta \tilde{k}^2)$ in Equation  \eqref{subsec:discuss:uniform_viscosity_DR} is independent on viscosity or Reynolds number $\mathcal{R}$.
Therefore, the critical wavelength remains unchanged when considering the effect of viscosity.
Figure \ref{fig:discuss:visc_1D}, which shows the dispersion relation including viscosity with $\alpha = 0.5,\ \beta = 0.01$.
The growth rates are suppressed by the viscosity but still positive. 
The unstable region $k < k_{\rm crit} \simeq 7k_n$ is not affected.

\begin{figure}
	\centering
	\includegraphics[width=0.5\linewidth]{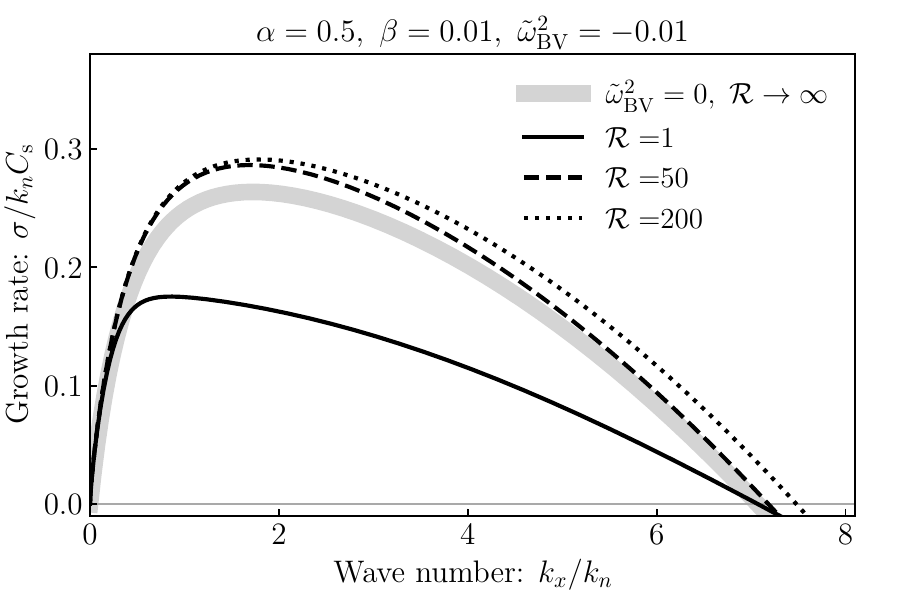}
	\caption{The dispersion relation including viscosity in the gravitationally stratified medium for $\alpha = 0.5$, $\beta = 0.01$, $\omega_{\mathrm{BV}}^2 = -0.01$, and $k_z = 0$. 
	The discrepancies among the black lines originate from differences in the Reynolds number $\mathcal{R} = 1, 50, 200$, 
	while the gray solid line represents the dispersion relation in the uniform medium without considering viscosity.}
	\label{fig:discuss:visc_2D}
\end{figure}

\paragraph*{In gravitationally stratified medium: }
Next, we extend the above discussion to a gravitationally stratified medium.
Since Unperturbed state in this case is also stationary, replacing $\sigma$ with $\sigma + \nu k^2$ in Equations \eqref{eq:2D:Linear_EoMx} and \eqref{eq:2D:Linear_EoMz}.
This substitution is equivalent to modifying the momentum equation to $\sigma \bar{m} n_0 \delta \bm{v} = -i\bm{k} \delta P - \bar{m} n_0 \nu k^2 \delta \bm{v} - \bar{m}g\bm{e}_z \delta n$, and reanalyzing yields the following dispersion relation, similar to the analysis for uniform medium,
\begin{align}
	(\tilde{\sigma} + \mathcal{R}^{-1} \tilde{k}^2) [ \tilde{\sigma}^3 + \{ \alpha + (\beta + \mathcal{R}^{-1}) \tilde{k}^2 \} \tilde{\sigma}^2
	+ \{ 1 + \mathcal{R}^{-1} (\alpha + \beta \tilde{k}^2) \} \tilde{k}^2 \tilde{\sigma} - \gamma^{-1} \tilde{k}^2 (1 - \alpha - \beta \tilde{k}^2) ] - |\tilde{\omega}_{\rm BV}^2| \tilde{k}_x^2 = 0.
	\label{subsec:discuss:viscosity_DR} 
\end{align}
Here, we replace $\tilde{\omega}_{\rm BV}^2$ with $-|\tilde{\omega}_{\rm BV}^2|$ because $\tilde{\omega}_{\rm BV}^2 < 0$ in the convectively unstable case.
The term in brackets in Equation \eqref{subsec:discuss:viscosity_DR} is identical to the dispersion relation given in Equation \eqref{subsec:discuss:uniform_viscosity_DR}.
Figure \ref{fig:discuss:visc_2D} shows the dispersion relation including viscosity in the gravitationally stratified medium. 
When the Reynolds number is small, i.e., $\mathcal{R} = 1$, the growth rate is reduced.
Conversely, for the large number ($\mathcal{R} = 200$), the growth rate becomes larger compared to the uniform medium case without viscosity.
Then, we focus on the critical wavelength $\lambda_{\rm crit} = 2\pi / k_{\rm crit}$ of the marginal state ($\sigma = 0$).
Figure \ref{fig:discuss:visc_2D} indicates that the critical wavelength is identical to the Field length (see Section \ref{subsec:1D_crit}) when the Reynolds number is small, i.e., $\mathcal{R} = 1$.
When $\mathcal{R} = 200$, the critical wavelength is smaller than the Field length ($\lambda_{\rm crit} < \lambda_{\rm F}$).
This discrepancy arises from the difference in the critical wavelengths for thermal and convective instabilities.
The critical wave number of convective instability, $k_{\rm crit, CI}$, can be estimated from the term in the dispersion relation that is independent on the growth rate.
This term is given by $-\gamma^{-1}\mathcal{R}^{-1}\tilde{k}^4 (1-\alpha - \beta \tilde{k}^2) - |\tilde{\omega}_{\rm BV}^2| \tilde{k}_x^2$.
To isolate the convective instability, we neglect the effect of radiative cooling by considering the limit $(1-\alpha) \to 0$.
This simplifies the term to $\gamma^{-1}\mathcal{R}^{-1}\beta \tilde{k}^6 - |\tilde{\omega}_{\rm BV}^2| \tilde{k}_x^2$.
The wave number for marginal condition ($\sigma = 0$) is found by setting this expression to zero, which yields:
\begin{align}
	\gamma^{-1} \mathcal{R}^{-1} \beta \tilde{k}_{\rm crit, CI}^6 - | \tilde{\omega}_{\rm BV}^2 | \tilde{k}_{x, {\rm crit, CI}}^2 = 0,
	\label{fig:discuss:visc_CIcrit1}
\end{align}
which is defined by both viscosity ($\mathcal{R}$) and thermal conduction ($\beta$) (see, \citealp{Seno_Inutsuka2025conv}, for more detail).
For perturbations with $k_z = 0$ (i.e., $k = k_x$), this expression can be solved to find the critical wave number:
\begin{align}
	\tilde{k}_{x, {\rm crit, CI}}^4 = \dfrac{\gamma | \tilde{\omega}_{\rm BV}^2 |}{\mathcal{R}^{-1}\beta}\quad {\rm or}\quad  \tilde{\lambda}_{x, {\rm crit, CI}} = 2\pi\left [ \dfrac{\mathcal{R}^{-1}\beta}{\gamma | \tilde{\omega}_{\rm BV}^2 |} \right ]^{1/4}.
	\label{fig:discuss:visc_CIcrit2}
\end{align}
In the case of $\alpha = 0.5,\ \beta = 0.01,\ {\rm and}\ \tilde{\omega}_{\rm BV}^2 = -0.01$, the critical wave number of thermal instability is $\tilde{k}_{\rm crit, TI} \simeq 7$.
Thus, the overall critical wave number is defined by the larger of the two wave numbers (corresponding to the smaller wavelength):
\begin{align}
	k_{\rm crit} = \mathrm{max} ( k_{\rm crit, TI},\ k_{\rm crit, TI} ) \quad {\rm or} \quad \lambda_{\rm crit} = \mathrm{min} ( \lambda_{\rm F},\ \lambda_{\rm crit, CI} ) . 
\end{align}
This indicates that we must consider the critical wavelength arising from both thermal and convective instabilities.

\subsection{Convectively stable case ($\omega^2_{\rm BV} > 0$)}
\label{subsec:discuss:CIn}

\begin{figure}
	\centering
	\includegraphics[width=1.0\linewidth]{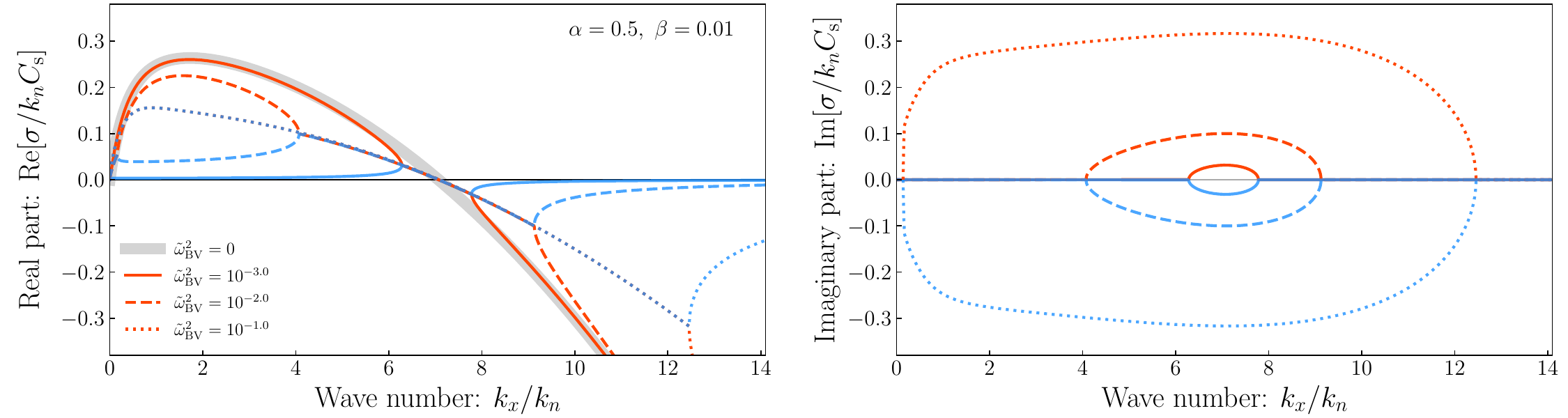}
	\caption{
	Dispersion relation in the convectively stable case with parameters $\alpha = 0.5$, $\beta = 0.01$, $\gamma = 5/3$, and $k_z = 0$.  
	The left panel shows the real part of the eigenvalues (the growth rate), while the right panel shows the imaginary part (the oscillation frequencies).
	The red lines correspond to thermally unstable modes in the uniform case, while the blue curves represent lines that originate from vertical oscillations governed by Equation~\eqref{eq:2D:Linear_EoMz}.
	}
	\label{fig:discuss_DR:Cn}
\end{figure}

We now examine the convectively stable case ($\omega^2_{\rm BV} > 0$).
In this case,  the growth rate $\sigma$ has both the real part (growth or decaying) and imaginary part (oscillation). 
A positive real part with an imaginary part indicates an over-stable mode in which the perturbation grows with oscillating.
Note that we neglect the effects of viscosity here.
While viscosity is one of the important effects to stabilize convective instability, as discussed in Section \ref{subsubsec:discuss:viscosity}, it is omitted here because we are considering the convectively stable case where $\omega_{\rm BV}^2 > 0$.
Figure~\ref{fig:discuss_DR:Cn} shows the dispersion relation obtained by solving Equation~\eqref{eq:2D:DR_WKB2}.  
The left panel displays the real part of the growth rate, while the right panel shows the imaginary part (oscillation frequencies).  
In this case, the thermal instability is suppressed by the effect of buoyancy, as indicated by the red lines in the left panel.  
On the other hand, the critical wavelength remains at the classical Field length of $\lambda_{\rm F} = \sqrt{\kappa |\partial \mathcal{L} / \partial T|_P^{-1}}$.
The buoyancy force only reduces the growth rate but does not eliminate the instability.

\subsection{Approximate solutions}
\label{subsec:2D:Approx}

In this section, we derive approximate forms of the dispersion relation in the short- and long-wavelength limits, following the procedure in Sections~\ref{subsec:1D_short} and~\ref{subsec:1D_long}.

\subsubsection{Short-wavelength approximation}
Under the short wavelength approximation, $\delta P = 0$, the dispersion relation reduces to
\begin{align}
	\tilde{\sigma} [ \tilde{\sigma} - \gamma^{-1} (1 - \alpha - \beta \tilde{k}^2) ] + \tilde{\omega}_{\rm BV}^2 = 0.
	\label{eq:2D:DR_short}
\end{align}
This expression is consistent with previous analyses that employed the incompressible or Boussinesq approximations (e.g., \citealp{Defouw1970, Binney_etal2009, Nipoti2010, McCourt_etal2012}).
As in Section~\ref{subsec:2D:DR}, the dispersion relation for a uniform medium appears explicitly in this expression.
The two solutions for the growth rate are given analytically by
\begin{align}
	&
	\tilde{\sigma}_\pm = \dfrac12 \left [ \tilde{\sigma}_{\rm TI} (k) \pm  \sqrt{ \tilde{\sigma}_{\rm TI}^2 (k) - 4 \tilde{\omega}_{\rm BV}^2 } \right ],
	\label{eq:2D:DR_short:solution}\\[1mm]
	&
	\tilde{\sigma}_{\rm TI} (k) \equiv \gamma^{-1} (1 - \alpha - \beta \tilde{k}^2),
	\label{eq:2D:DR_short:solution_normTI}
\end{align}
where $\tilde{\sigma}_{\mathrm{TI}}$ represents the dimensionless growth rate of thermal instability at a given wave number $k$.
Note that $\tilde{\sigma}_{\mathrm{TI}} (k)$ is the same as the dispersion relation in uniform medium under the short-wavelength approximation (see Equation \eqref{eq:1D:Linear_DR:short}).
When $\tilde{\sigma}_{\mathrm{TI}} (k) > 0$, the mode $\tilde{\sigma}_+$ corresponds to the classical thermally unstable mode.

As Figure \ref{fig:discuss_DR:Cn} suggests, both sets of modes can exhibit over-stability ($\sigma \in \mathbb{C}$) and corresponding range of wave number becomes wider with a stronger gravity ($\omega_{\rm BV}^2\propto g$). 
From the reduced dispersion relation, Equation \eqref{eq:2D:DR_short:solution}, the comparison of characteristic rates of thermal instability $\sigma_{\rm TI}$ and buoyant force oscillation $4\omega_{\rm BV}^2$ determines whether the over-stable mode ($\sigma \in \mathbb{C}$) appears.
When the subtraction $\sigma_{\rm TI}^2 - 4 \omega_{\rm BV}^2 > 0$, the growth rate is real and purely exponential ($\sigma \in \mathbb{R}$).  
When $\sigma_{\mathrm{TI}}^2(k) - 4\omega_{\mathrm{BV}}^2 < 0$, the over-stable mode appear ($\sigma \in \mathbb{C}$).
We defer detailed discussion of their implications to a later section.

\subsubsection{Long-wavelength approximation}
Under the long wavelength approximation, which any deviations from the unperturbed state due to internal energy (i.e., entropy) fluctuations are rapidly adjusted to the equilibrium state.
This reduces the linearized energy equation \eqref{eq:2D:Linear_Energy} to the following form:
\begin{align}
	\frac{1}{\gamma - 1} \frac{dP_0}{dz} \delta v_z  -  \frac{\gamma P_0}{(\gamma - 1)n_0} \frac{dn_0}{dz} \delta v_z
      	=  - \mathcal{L}_n \delta n - \mathcal{L}_T \delta T  - \overline{m} n_0 g \delta v_z.
	\label{eq:2D:DR_longApprox}
\end{align}
By substituting this result into the continuity equation \eqref{eq:2D:Linear_EoC} and momentum equations \eqref{eq:2D:Linear_EoMx} and \eqref{eq:2D:Linear_EoMz}, we derive the following approximate form for the real part of the dispersion relation:
\begin{align}
	\tilde{\sigma} [ \alpha \tilde{\sigma} ^2 - \gamma^{-1} \tilde{k}^2 ( 1 - \alpha ) ] + \tilde{\omega}_{\rm BV}^2  \tilde{k}_x^2 = 0.
	\label{eq:2D:DR_long}
\end{align}
As with previous derivations of the dispersion relations, we confirm that the expression within the brackets agrees exactly with that in uniform medium under the long-wavelength approximation $\alpha \tilde{\sigma}^2 - \gamma^{-1}(1-\alpha)\tilde{k}^2 = 0$.

\subsection{Estimation of the most unstable condition}
\label{subsec:discuss:MUC}

We provide approximate estimates for the most unstable wavelength and growth rate in stratified systems.
We distinguish between two cases: (i) purely unstable modes, for which $\sigma \in \mathbb{R}$, and (ii) completely over-stable modes, for which $\sigma \in \mathbb{C}$ at all wavelengths.
Purely unstable modes include both the convectively unstable ($\omega_{\rm BV}^2 < 0$) and the convectively stable ($\omega_{\rm BV}^2 > 0$) cases where $\sigma_{\mathrm{TI}}^2(k) - 4\omega_{\mathrm{BV}}^2 < 0$.

\subsubsection{Purely unstable case}

We begin by considering the purely unstable case, which includes both the convectively unstable case and the convectively stable case where $\tau_{\mathrm{TI}}(k) < \tau_{\mathrm{BV}}/2$.
It is generally difficult to estimate the most unstable mode in the stratified case in the same manner as for the uniform medium (Section~\ref{subsec:1D_MostUnstable}).  
Instead, we employ a simplified approach based on the short-wavelength approximation.

Assuming $\sigma_{\mathrm{TI}} (k)$ is much greater than $\omega_{\mathrm{BV}}$, i.e., $\sigma_{\mathrm{TI}} (k) \gg \omega_{\mathrm{BV}}$, we use Equation~\eqref{eq:2D:DR_short:solution} to expand the larger and smaller roots of the dispersion relation as:
\begin{align}
	\sigma_+ &\approx \left[ 1 - \left( \frac{\omega_{\mathrm{BV}}}{\sigma_{\mathrm{TI}}} \right)^2 \right] \sigma_{\mathrm{TI}},
	\label{eq:2D:discuss:MUC_pGrowth} \\[1mm]
	\sigma_- &\approx \left( \frac{\omega_{\mathrm{BV}}}{\sigma_{\mathrm{TI}}} \right)^2 \sigma_{\mathrm{TI}}.
	\label{eq:2D:discuss:MUC_mGrowth}
\end{align}
Assuming that Equation~\eqref{eq:2D:discuss:MUC_pGrowth} holds at the most unstable point, we identify $\sigma_{+,\mathrm{max}}$ as the maximum growth rate.  
We then substitute this into the expressions from Section~\ref{subsec:1D_MostUnstable} to estimate the corresponding wavenumber.
Note that $\sigma_-$ is stable when $\omega_{\mathrm{BV}}^2 < 0$ and exhibits no peak in the purely unstable regime when $\omega_{\mathrm{BV}}^2 > 0$, so we focus only on the $\sigma_+$ branch.  
The corresponding most unstable wavenumber is obtained by substituting $\sigma_{+,\mathrm{max}}$ into either Equation~\eqref{eq:1D:Linear_DR:most_derivative1} or~\eqref{eq:1D:Linear_DR:most_derivative2}:
\begin{align}
	\tilde{k}_{\mathrm{max}}^2 \approx \left[ 1 - \left( \frac{\omega_{\mathrm{BV}}}{\sigma_{\mathrm{TI}}} \right)^2 \right] \tilde{k}_{\mathrm{max,TI}}^2 
	= \left[ 1 - \left( \frac{\omega_{\mathrm{BV}}}{\sigma_{\mathrm{TI}}} \right)^2 \right]
	\left[ 2 + \sqrt{ \frac{\gamma}{|\alpha| \beta} } \right]^{-1} \tilde{k}_{\mathrm{crit}}^2.
	\label{eq:2D:discuss:MUC_WN}
\end{align}
Note that this estimation is valid only when $\sigma_{\mathrm{TI}}(k) > 2\omega_{\mathrm{BV}}$.

\subsubsection{Completely over-stable case}

We turn to the case of complete over-stability (Figure \ref{fig:discuss_DR:SchematicOS}), in which the growth rate $\sigma$ is complex at all wavelengths, except in the limits $k \to 0$ or $k = k_{\mathrm{crit}}$.  
In this case, since $\omega_{\rm BV}^2 > 0$, the real part of the growth rate $\mathrm{Re}[\sigma]$ is determined by the growth rate of thermal instability in uniform medium.
From Equation~\eqref{eq:2D:DR_short:solution}, we note that the magnitude of the growth rate is half that of the purely thermal instability:
\begin{align}
	\mathrm{Re}[\sigma] = \frac{1}{2} \sigma_{\mathrm{TI}}.
\end{align}
Assuming that this holds at the most unstable point, we estimate the maximum growth rate as
\begin{align}
	\mathrm{Re}[\sigma]_{\mathrm{max}} \approx \frac{1}{2} \sigma_{\mathrm{max,TI}}.
\end{align}
As Equation \eqref{eq:2D:DR_short:solution} suggests, when $\omega_{\rm BV}^2 > 0$, the buoyancy force can only reduce the growth rate of thermal instability by a factor of $\approx 0.5$. 
Therefore, a significant extension of the cooling time is not expected.
Furthermore, we obtain the corresponding wave number by substituting into Equation~\eqref{eq:1D:Linear_DR:most_derivative2}, yielding
\begin{align}
	k_{\mathrm{max}} \approx \frac{1}{\sqrt{2}} k_{\mathrm{max,TI}}.
\end{align}
This provides a convenient scaling relation for estimating the most unstable conditions in this case.

\section{Discussions} \label{sec:discuss}
In this section, we examine in detail how thermal instability is modified by convection, using the results derived in Section~\ref{sec:2D}.  
First, we perform a linear analysis in the Lagrangian coordinates to qualitatively discuss the motion of fluid elements in Section \ref{subsec:discuss:Lagrange}.
We examine both the convectively unstable ($\omega_{\rm BV}^2 < 0$) and the convectively stable ($\omega_{\rm BV}^2 < 0$) cases.  
For simplicity, we consider modes with a vertical wave number of $k_z = 0$, so that all perturbations lie in the horizontal plane.
Then, in Section \ref{subsec:discuss:Estimation}, we extend our estimation to the simplified CGM model with a linear temperature gradient. 
We then derive the typical time and length scales associated with the formation of high-velocity cloud (HVC).

\subsection{Vertical motion of a fluid element in the Lagrangian coordinates}
\label{subsec:discuss:Lagrange}
In this section, we perform a linear analysis in Lagrangian coordinates to explicitly discuss the vertical motion of a fluid element.
Focusing on the short-wavelength approximation allows us to obtain a clear, qualitative understanding of how thermal instability is affected by gravity, an approach that is difficult to grasp from the more complex general dispersion relation.

We introduce the vertical displacement, $\xi_z$, where $d\xi_z/dt= \delta v_z$  and we define $\dot{\xi} \equiv \sigma \xi_z$. 
The horizontal displacement $\xi_x$ is initially zero, but the fluid element moves to the $x$-direction by its pressure gradient according to the equation of motion \eqref{eq:2D:Linear_EoMx}.
Thus, the trajectory indicates a vortex motion.
For simplicity and to focus on the gravity effect on thermal instability, we only discuss the vertical motion under the isobaric approximation, as described by Equation \eqref{eq:2D:DR_short:solution_normTI}.

Under the short wavelength approximation ($\delta P = 0$), the linearized energy equation \eqref{eq:2D:Unp_Energy} simplifies, and by substituting $\delta v_z = \sigma \xi_z$, the vertical motion of a fluid element is governed by the following equation:
\begin{align}
	\ddot{\xi}_z - \sigma_{\rm TI} (k)  \dot{\xi}_z + \omega_{\rm BV}^2 \xi_z = 0.
	\label{eq:2D:Lagrange_EoMshort}
\end{align}
In the thermally unstable case $\sigma_{\mathrm{TI}}(k) > 0$, the second term $\sigma_{\rm TI} (k)  \dot{\xi}_z$ represents the runaway acceleration in the vertical $z$-direction, because the thermal instability enhances the buoyancy force. 
On the other hand, the third term $\omega_{\rm BV}^2 \xi_z$ indicates the stabilizing resilience when $\omega_{\mathrm{BV}}^2 > 0$, while it acts as the driving force of convection in the case of $\omega_{\mathrm{BV}}^2 < 0$.
This analysis describes the effects of buoyancy and thermal instability on the evolution of vertical fluid displacements. 
This reveals two distinct regimes: one where thermal instability is stabilized by buoyant resilience forces (for $\omega_{\mathrm{BV}}^2 > 0$), and another where it is enhanced by convection (for $\omega_{\mathrm{BV}}^2 < 0$). 
The detailed picture is discussed in Sections \ref{subsubsec:discuss:CyLagrange} and \ref{subsubsec:discuss:CnOS}.

\subsubsection{Convectively unstable case ($\omega_{\rm BV}^2 < 0$)}
\label{subsubsec:discuss:CyLagrange}
In this section, we qualitatively discuss the reason why the thermal instability is enhanced in convectively unstable case ($\omega_{\rm BV}^2 < 0$).
We denote the larger and smaller growth rates as $\sigma_+$ and $\sigma_-$, respectively, following the convention in Equation~\eqref{eq:2D:DR_short:solution}.  
The corresponding vertical displacements are denoted as $\xi_z^+$ and $\xi_z^-$.
We rewrite Equation \eqref{eq:2D:Lagrange_EoMshort} as $\ddot{\xi}_z = \sigma_{\mathrm{TI}} (k) \dot{\xi}_z + |\omega_{\mathrm{BV}}^2| \xi_z $.
Here, we replace $\tilde{\omega}_{\rm BV}^2$ with $-|\tilde{\omega}_{\rm BV}^2|$ because $\tilde{\omega}_{\rm BV}^2 < 0$ in this case.
The displacements $\xi_z^\pm$ evolve as $\xi_z^\pm \propto \exp(\sigma_\pm t)$, with $\sigma_+ > 0$ corresponding to an unstable (growing) mode and $\sigma_- < 0$ to a stable (decaying) mode.
Physically, $\xi_z^+$ describes a dense fluid element accelerating downward, whereas $\xi_z^-$ represents its upward motion.
This is because the buoyancy force acts downward on a dense fluid element, and thermal instability further enhances this downward acceleration (instability).
Conversely, for the upward motion ($\xi_z^-$), the buoyancy acts as a decelerating force, which is further enhanced by thermal instability, leading to stabilization.

\subsubsection{Convectively stable case ($\omega_{\rm BV}^2 > 0$)}
\label{subsubsec:discuss:CnOS}
We discuss the convectively stable case ($\omega^2_{\rm BV} > 0$).
We can rewrite Equation \eqref{eq:2D:Lagrange_EoMshort} as $\ddot{\xi}_z = \sigma_{\mathrm{TI}}(k) \dot{\xi}_z - \omega_{\mathrm{BV}}^2 \xi_z$.
Now that we are considering a thermally unstable fluid element, i.e. $\sigma_{\mathrm{TI}}(k) > 0$, the first term $\sigma_{\rm TI}\dot{\xi}_z$ describes a runaway acceleration.
The second term $-\omega_{\rm BV}^2 \xi_z$ describes a restoring force, and thus plays a stabilizing role against thermal instability, which results in a decrease of the growth rate.

We now turn our attention to the second family of modes, represented by the blue lines in Figure \ref{fig:discuss_DR:Cn}. 
These modes arise from the vertical motion governed by the $z$-component of the momentum equation, Equation~\eqref{eq:2D:Linear_EoMz}. 
They also become unstable in the convectively stable case ($\omega_{\rm BV}^2>0$). 
In contrast, in the convectively unstable case ($\omega_{\rm BV}^2<0$), where these modes remained stable.

\begin{figure}
	\centering
	\includegraphics[width=\linewidth]{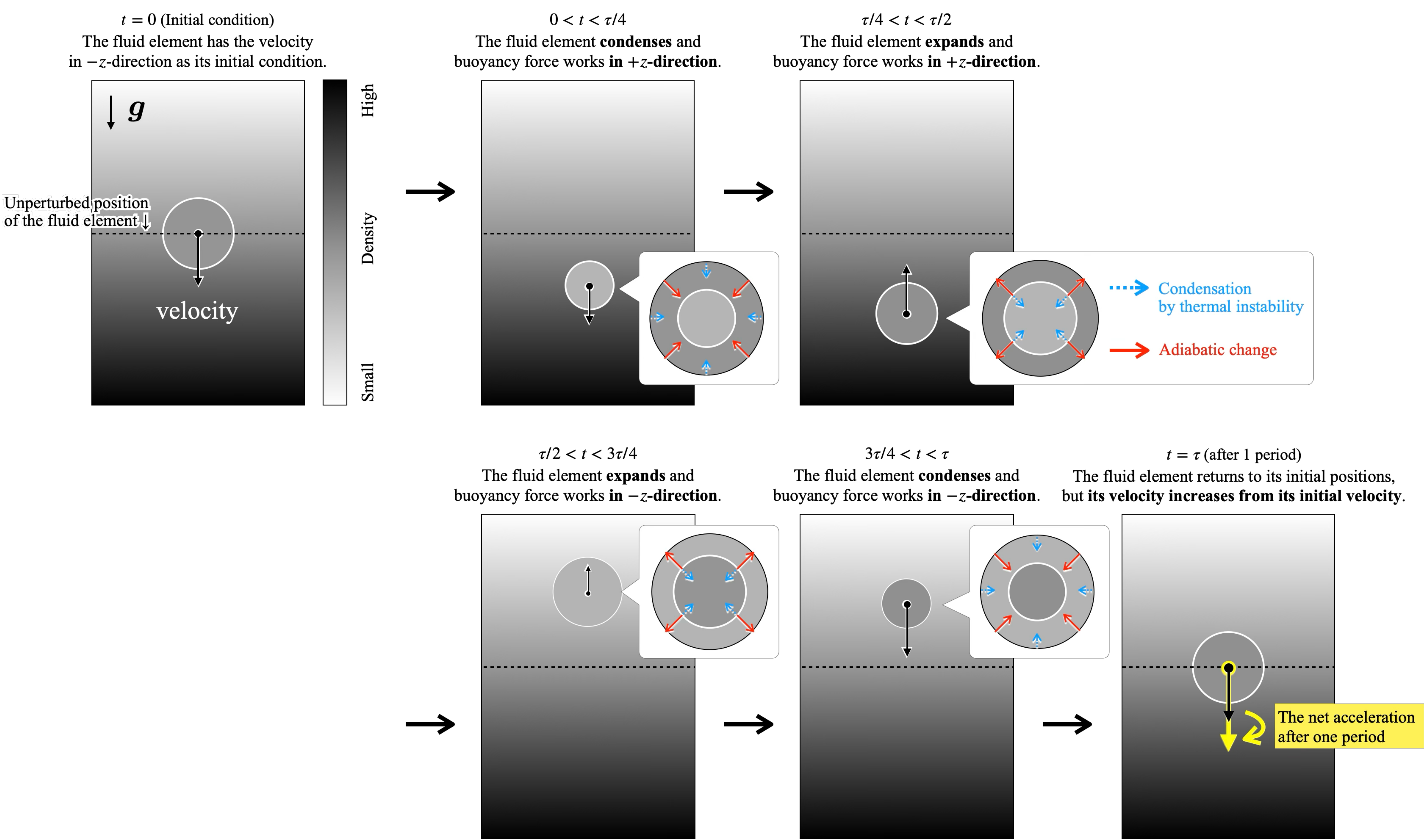}
	\caption{
	Schematic diagram illustrating the over-stable behavior of a thermally unstable fluid element in the $z$-direction, where $\sigma_{\rm TI} (k) < 2\omega_{\rm BV}$,  over one oscillation period ($0  \le t \le \tau$), during which the element oscillates and grows in amplitude.
	The circles indicate the fluid element, while the black and red arrows represent its velocity direction.
	Additionally, the yellow arrow represents the net displacement after one oscillation period,  $\tau \equiv 2\pi [\omega_{\rm BV}^2 - (\sigma_{\rm TI}(k)/2)^2]^{-1/2}$
	The background shading indicates the density distribution.
	}
	\label{fig:discuss_DR:SchematicOS}
\end{figure}
We discuss why the over-stable mode appears only in the convectively stable case ($\omega_{\rm BV}^2 > 0$).
We introduce the characteristic time scale of thermal instability $\tau_{\rm TI} = \sigma_{\rm TI}^{-1}$ and buoyant oscillation $\tau_{\rm BV}/2 = (2\omega_{\rm BV})^{-1}$.
Figure~\ref{fig:discuss_DR:SchematicOS} schematically illustrates the dynamics of a fluid element with the over-stable condition of $\tilde{\sigma}_{\rm TI}^2 (k) - 4\tilde{\omega}_{\rm BV}^2 < 0$. 
In this case, with $\tau_{\mathrm{TI}}(k) > \tau_{\mathrm{BV}}/2$, the fluid element is not sufficiently cooled during half of an oscillation period, allowing it to return to a location above its original position. 
A key finding here is that the factor of $1/2$ represents this half-period of oscillation.
The competition between buoyancy and cooling gives rise to over-stable oscillatory motion. 
The detailed behavior over one oscillation period, $\tau \equiv 2\pi [\omega_{\rm BV}^2 - (\sigma_{\rm TI}(k)/2)^2]^{-1/2}$, is as follows:
\begin{enumerate}
	\item $t = 0$ [Top-left]: \\
	We initiate the system with a thermally unstable fluid element that has zero vertical displacement ($\xi_z = 0$) and an initial downward velocity ($\dot{\xi}_z < 0$).
	Under this initial condition, the element's density matches that of its surroundings, resulting in zero buoyancy force.
	\item $0 < t < \tau / 4$ [Top-second]: \\
	As the element moves downward, it is displaced into a position of higher background pressure, leading to adiabatic compression until isobaric conditions are achieved. 
	This compression makes the element less dense than its surroundings. 
	Consequently, a buoyancy in the $+z$-direction acts on the fluid element, causing it to decelerate.
	\item $\tau / 4 < t < \tau / 2$ [Top-third]: \\
	After decelerating completely and reaching zero velocity at $t = \tau / 4$, the upward buoyancy causes the element to begin rising.
	As it rises, it undergoes adiabatic expansion.
	While this expansion opposes the condensation driven by thermal instability, the adiabatic effect dominates in this regime, i.e., $\tau_{\mathrm{TI}}(k) > \tau_{\mathrm{BV}}/2$, thereby maintaining the element's lower density.
	The element therefore continues to accelerate upward due to the sustained positive buoyancy. 
	The competition between opposing effects ensures the element expands less than it would under a pure adiabatic expansion.
	\item $\tau / 2 < t < 3\tau / 4$ [Bottom-left]: \\
	The element continues rising into a position of lower background pressure and undergoes adiabatic expansion. 
	This expansion causes the element to become denser than its surroundings in this phase. 
	Consequently, a negative buoyancy acts on the element, causing it to decelerate. 
	Similar to the previous expansion phase ($\tau / 4 < t < \tau / 2$), the thermal instability counteracts the adiabatic expansion, making the expansion less effective than a pure adiabatic change.
	\item $3\tau / 4 < t < \tau$ [Bottom-second]: \\
	After reaching zero velocity at $t=3\tau/4$, the downward buoyancy causes the element to begin falling. 
	As it falls, the element is compressed by the surrounding medium. 
	Critically, this adiabatic compression is aligned with the condensation due to thermal instability. 
	This cooperative action causes the element to be compressed more than it would be under pure adiabatic conditions.
	Consequently, the resulting negative buoyancy force accelerates the element's fall. 
	Note that if $\sigma_{\rm TI}=0$, the compression would be purely adiabatic, resulting only in a stable buoyant oscillation.
	\item $t = \tau$ [Bottom-third]: \\
	After one oscillation period, the fluid element has lost internal energy due to radiative cooling (thermal instability). 
	It now possesses a greater downward speed than its initial speed at $t=0$. 
	This energy loss fuels the instability, causing the element's oscillation amplitude to increase in the subsequent phase. 
	The element then enters the next period with a larger condensation rate due to its increased density and amplitude.
\end{enumerate}
Throughout this oscillation cycle, the condensation effect of thermal instability is crucial. 
It inhibits the expansion of fluid elements while promoting their compression. 
This net effect results in over-stable behavior, causing the fluid element to gradually fall, even in a convectively stable system that is otherwise characterized by buoyant oscillation.

\begin{figure}
	\centering
	\includegraphics[width=0.7\linewidth]{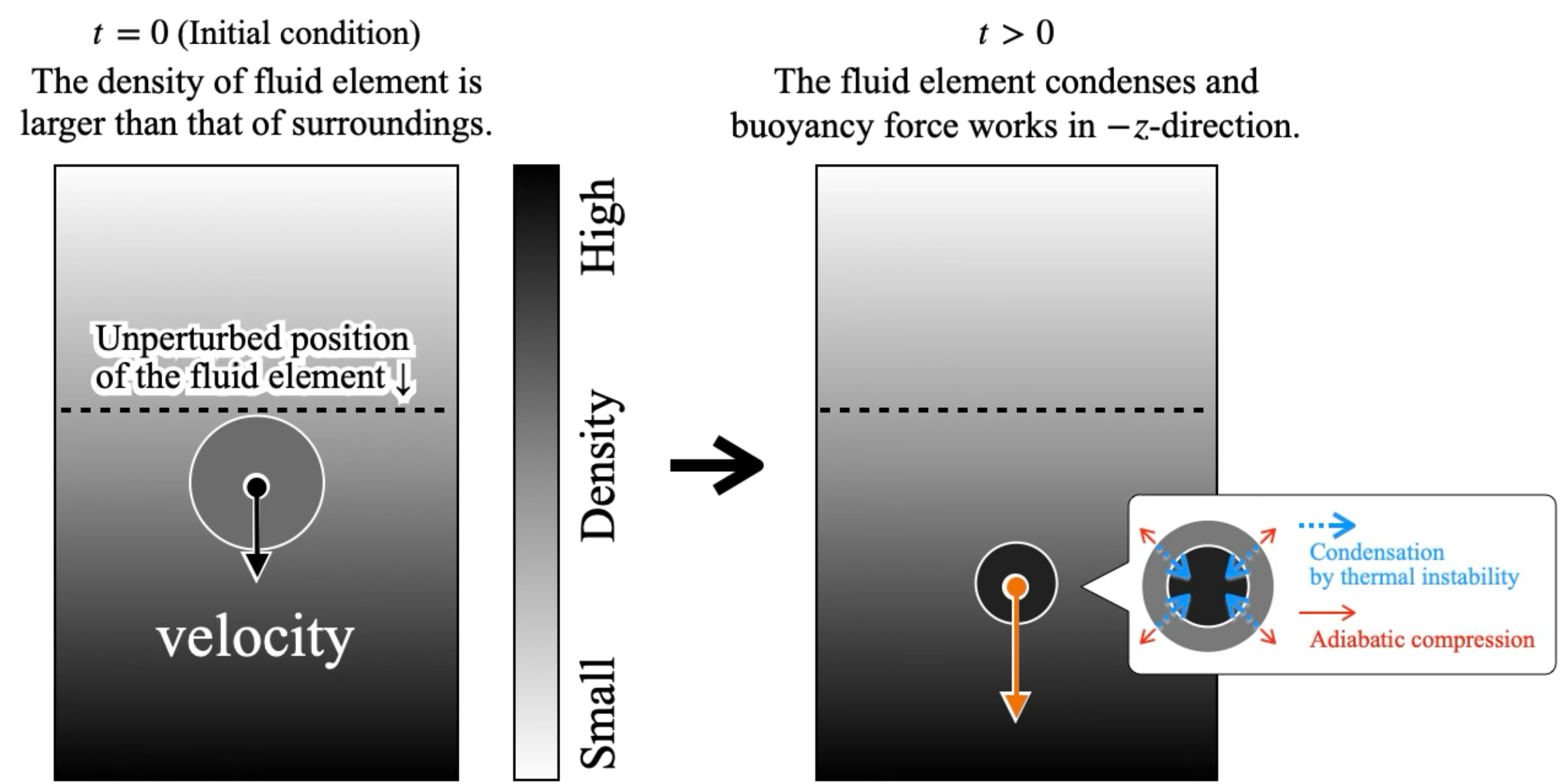}
	\caption{
	Schematic diagram illustrating the unstable behavior of a thermally unstable fluid element in the $z$-direction, where $\sigma_{\rm TI} (k) > 2\omega_{\rm BV}$.
	The circles indicate the fluid element, while the background shading indicates the density distribution.
	}
	\label{fig:discuss_DR:SchematicU}
\end{figure}
In contrast, when the system is purely unstable, i.e., $\tau_{\mathrm{TI}}(k) < \tau_{\mathrm{BV}}/2$, the fluid element is sufficiently cooled while descending, as illustrated in Figure \ref{fig:discuss_DR:SchematicU}. 
This sufficient cooling causes the element to reach a denser region below its original position, where it is further compressed by the surrounding pressure. 
This process leads to the monotonic growth of the instability.

\subsection{Simple estimation for the equilibrium CGM}
\label{subsec:discuss:Estimation}
In this section, we briefly estimate the most unstable wavelength and its growth rate for typical CGM parameters, using Equations \eqref{eq:2D:discuss:MUC_pGrowth} and \eqref{eq:2D:discuss:MUC_WN}.
Specifically, we adopt the following typical unperturbed physical parameters for the CGM: the number density $n_0 = 10^{-3}\ {\rm cm}^{-3}$ and temperature $T_0 = 10^6\ {\rm K}$.
This is because, by applying a total hot halo gas mass of $10^{11}\ M_\odot$ which is estimated from X-ray observations (e.g., \citealp{Fang+2013, Miller_Bregman2013, Miller_Bregman2015, Nakashima_2018}), the mean number density and the virial temperature of the Milky Way for a distance from the galactic center of $\sim\ 300\ {\rm kpc}$ are derived as $n_{\rm mean}\sim 10^{-3}\ {\rm cm}^{-3} (\mu / 0.6)^{-1}(M_{\rm gas} / 10^{11} M_\odot) (r/300\ {\rm kpc})^{-3}$ and $T_{\rm vir} \sim 10^6\ {\rm K} (M_{\rm vir} / 10^{12} M_\odot)(r/300\ {\rm kpc})^{-1}$, respectively.
Using these parameters, we estimate the cooling time and Field length as
\begin{align}
	&
	t_{\rm cool} \sim \dfrac{k_{\rm B}T_0}{n_0 \mathit\Lambda} \approx 4.6 \times 10^1 \ {\rm Myr}\ 
	\left (  \dfrac{n_0}{10^{-3}\ {\rm cm}^{-3}} \right )^{-1} \left (  \dfrac{T_0}{10^6\ {\rm K}} \right ) \left ( \dfrac{\mathit\Lambda}{10^{-22}\ {\rm erg\ s^{-1}\ cm^3}} \right )^{-1},
	\label{eq:discuss:Estimation_CoolingTime} \\[3mm]
	&
	\lambda_{\rm F} \sim 2\pi \sqrt{ \dfrac{\kappa T_0}{n_0^2 \mathit\Lambda}} 
	\approx 8.4\ {\rm kpc}\ \left ( \dfrac{n_0}{10^{-3}\ {\rm cm}^{-3}} \right )^{-1} 
	\left ( \dfrac{T_0}{10^6\ {\rm K}} \right )^{7/4}\left ( \dfrac{\mathit\Lambda}{10^{-22}\ {\rm erg\ s^{-1}\ cm^3}} \right )^{-1/2},
	\label{eq:discuss:Estimation_FieldLength}
\end{align}
where we use the thermal diffusivity for fully-ionized plasma $\kappa (T) = 1.6 \times 10^{-6} (T/{\rm K})^{5/2}\ {\rm erg\ K^{-1}cm^{-1}s^{-1}}$ (\citealp{Spitzer1963}) and a cooling function value for solar metallicity $\mathit\Lambda \sim 10^{-22}\ {\rm erg\ s^{-1}\ cm^3}$ (c.f., \citealp{Sutherland_Dopita1993, Schmutzler_Tscharnuter1993, Gnat2017}).
In addition, we assume a linear temperature gradient from $T = 10^6\ {\rm K}$ to $10^7\ {\rm K}$ throughout the virial radius of the Milky Way, $r_{\rm vir} \sim 300\ {\rm kpc}$.
The reason for assumption a positive temperature gradient is that warm HI occupies more volume at lower $z$-heights from the disk, while hot gas dominates at the higher $z$-heights (e.g., \citealp{Gaensler+2008, Savage_Wakker2009, Putman_etal2012b}).
Under this condition, the system is convectively stable ($\omega_{\rm BV}^2 > 0$), so this system can exhibit both unstable and over-stable modes and the critical wavelength corresponds to the Field length, as given by Equation \eqref{eq:discuss:Estimation_FieldLength} and discussed in Section \ref{subsec:discuss:CIn}.
Since we consider hydrostatic equilibrium, the inverse pressure scale height is given by $H_P^{-1} = \gamma g /  C_{\rm s}^2$.
The $Brunt-V\ddot{a}is\ddot{a}l\ddot{a}$ frequency is then estimated to be $\omega_{\rm BV} \approx 4.4 \times 10^1\ {\rm Gyr}^{-1}$.
Using the inverse of the cooling time, as indicated in Equation \eqref{eq:discuss:Estimation_CoolingTime}, as the thermal instability growth rate, i.e., $\sigma_{\rm TI} \sim t_{\rm cool}^{-1}$, we derive the ratio of the $Brunt-V\ddot{a}is\ddot{a}l\ddot{a}$ frequency to the thermal instability growth rate as $\omega_{\rm BV} / \sigma_{\rm TI} \approx 0.20$ .

Using Equations \eqref{eq:1D:Linear_DR:most_wl} and \eqref{eq:2D:discuss:MUC_WN}, we estimate the most unstable wavelengths of thermal instability and the gravity-modified one as follows:
\begin{align}
 	& \lambda_{\rm max, TI} \sim \lambda_{\rm F} \left[ 2 + \dfrac{n_0 k_{\rm B} C_{\rm s}}{\gamma - 1} \sqrt{ \dfrac{\gamma T_0}{\kappa n_0^2 \mathit\Lambda} } \right]^{1/2} \approx 3.5 \times \lambda_{\rm F}\ \approx\ 2.9 \times 10^1\ {\rm kpc} \left ( \dfrac{\lambda_{\rm F}}{8.4\ {\rm kpc}} \right ),
	\label{eq:discuss:Estimation_maxTI}
	\\[1mm]
	& \lambda_{\rm max} \sim  \lambda_{\rm max, TI} \left[ 1 - \left( \frac{\omega_{\mathrm{BV}}}{\sigma_{\mathrm{TI}}} \right)^2 \right]^{-1} \approx 1.1 \times \lambda_{\rm max, TI}\ 
	\approx\ 3.2 \times 10^1\ {\rm kpc}  \left ( \dfrac{\lambda_{\rm F}}{8.4\ {\rm kpc}} \right ). 
	\label{eq:discuss:Estimation_maxTCI}
\end{align}
According to Equation \eqref{eq:discuss:Estimation_maxTCI}, the most unstable wavelength is found to be comparable to that of the uniform medium case ($\lambda_{\rm max, TI}$).
This allows us to determine the most unstable condition in this system by simply estimating the result from the uniform medium analysis (\citealp{Field1965}).

Some numerical simulations say that the thermal instability of the circum-galactic medium formes high-velocity clouds (e.g., \citealp{Ramesh_etal2024, Lucchini+2024, Lucchini+2025}).
High-velocity clouds is considered to be the fuel for star formation in the galactic disk.
Therefore, estimating their typical mass and size is crucial to examine their contribution to the star formation rate.
Given information on the galactic hot gas, specifically a typical density of $n_0 = 10^{-3}\ {\rm cm}^{-3}$, and the most unstable wavelength as $\lambda_{\rm max} \approx 32\ {\rm kpc}$, we can estimate their typical mass as follows:
\begin{align}
	M_{\rm HVC} \approx 8.5 \times 10^7\ M_\odot\ \left ( \dfrac{\mu}{0.6} \right ) \left ( \dfrac{n_0}{10^{-3}\ {\rm cm}^{-3}} \right ) \left ( \dfrac{\lambda_{\rm max}}{32\ {\rm kpc}} \right )^3 .
	\label{eq:discuss:Estimation_Mhvc}
\end{align}
Additionally, by assuming a typical HVCs temperature of $\sim 10^4\ {\rm K}$ (e.g., \citealp{Putman_etal2012b, Muller+1963}) and considering spherical and condensation by thermal instability, we estimate their typical density to be $n_{\rm HVC}\sim 10^{-1}\ {\rm cm}^{-3}$.
Thus, assuming the total mass of HVC is completely conserved (neglecting the ionization, evaporation, and so on), their typical size can be estimated as
\begin{align}
	r_{\rm HVC} \sim \left ( \dfrac{n_0}{n_{\rm HVC}} \right )^{1/3} \lambda_{\rm max}\ 
	\approx \ 6.9\ {\rm kpc} \left ( \dfrac{n_0}{10^{-3}\ {\rm cm}^{-3}} \right )^{1/3} \left (  \dfrac{n_{\rm HVC}}{10^{-1}\ {\rm cm}^{-3}} \right )^{-1/3} 
	\left ( \dfrac{\lambda_{\rm max}}{32\ {\rm kpc}} \right ) .
	\label{eq:discuss:Estimation_Lhvc}
\end{align}
Based on this estimation, simulations of HVC formation should resolve scales of at least $\sim 7$ kpc.

Furthermore, we estimate the most unstable growth time, $t_{\rm min}$, using Equations \eqref{eq:2D:discuss:MUC_pGrowth} and \eqref{eq:discuss:Estimation_CoolingTime} as follows:
\begin{align}
	t_{\rm min} \sim \sigma_{+, {\rm max}}^{-1} \sim  t_{\rm cool} \left[ 1 - \left( \frac{\omega_{\mathrm{BV}}}{\sigma_{\mathrm{TI}}} \right)^2 \right]^{-1} 
	\approx 1.1 \times t_{\rm cool}\ 
	\approx\ 5.1\times 10^1\ {\rm Myr} \left ( \dfrac{t_{\rm cool}}{46\ {\rm Myr}}\right ).
	\label{eq:discuss:Estimation_Tmin}
\end{align}
Since this growth time is much shorter than the dynamical time of $t_{\rm dyn} \approx 2.0\ {\rm Gyr}\ (r_{\rm vir}/300\ {\rm kpc})(C_{\rm s}/150\ {\rm km}\ {\rm s}^{-1})^{-1}$ in this system, this thermal instability-driven process must be taken into account.
Additionally, using Equations \eqref{eq:discuss:Estimation_Mhvc} and \eqref{eq:discuss:Estimation_Tmin}, we can estimate the contribution of HVCs formed from the galactic hot gas to the gas supply to the galactic disk as follows:
 \begin{align}
 	\dot{M}_{\rm HVC} \sim \dfrac{M_{\rm HVC}}{t_{\rm min}} \approx 1.7\ M_\odot\ {\rm yr}^{-1}\ \left ( \dfrac{M_{\rm HVC}}{8.5\times 10^7\ M_\odot} \right ) \left ( \dfrac{t_{\rm min}}{51\ {\rm Myr}} \right )^{-1}.
	\label{eq:discuss:Estimation_Min}
 \end{align}
 Since the current star formation rate is $\sim 3\ M_\odot\ {\rm yr}^{-1}$, our simple estimation suggests HVCs formed from the CGM may contribute in sustaining the star formation rate over the depletion time of the disk gas.

\section{Summary}
\label{sec:summary}

Thermal instability plays a fundamental role in shaping multi-phase structures in a variety of astrophysical environments. 
While many previous studies have analyzed thermal instability under the assumption of a uniform background, gravitational effects is important and should be investigated in large-scale systems such as the circum-galactic medium (CGM) and intracluster medium. 
In this study, we examine thermal instability in a gravitationally stratified medium, with particular focus on its interaction with convective stability.

We first derive analytical approximations to the dispersion relation in a uniform medium, which serves as a reference case. 
We then extend this analysis to stratified systems using the Wentzel-Kramers-Brillouin approximation, where buoyancy is incorporated through the $Brunt-V\ddot{a}is\ddot{a}l\ddot{a}$ frequency, which enables us to explore the impact of the stability of convection and instability on the thermal instability.

The main results of this study are as follows:
\begin{itemize}
    \item In convectively unstable case ($\omega_{\mathrm{BV}}^2 < 0$), the thermal instability is enhanced by buoyancy force.  
    As a result, instability can occur even at wavelengths shorter than the classical Field length, since convective instability is not suppressed by thermal conduction alone.
    Moreover, when viscosity is considered, it has been found that the critical wavelength for this phenomenon is determined by the shorter of the critical wavelength for convection and the Field length, where the Field length represents the critical wavelength of thermal instability in uniform medium.
    
    \item In convectively stable case ($\omega_{\mathrm{BV}}^2 > 0$), the system can exhibit over-stability, in which perturbations grow oscillatory.  
    Furthermore, we find that two unstable modes exist in this regime, in contrast to the single unstable mode found in the convectively unstable case.  
    We attribute this difference to the structure of the eigenfunctions in the presence of buoyant restoring forces.

    \item The structure of the eigenfunctions, particularly the vertical velocity perturbation $\dot{\xi}_z$, differs qualitatively between convectively stable and unstable cases.  
    The phase and growth characteristics of the vertical displacement $\xi_z^\pm$ depend sensitively on the sign of $\omega_{\mathrm{BV}}^2$.

    \item We propose approximate expressions for the most unstable growth rate and wavelength in gravitationally-stratified medium, extending the results of the uniform medium case by incorporating gravitational effects through the $Brunt-V\ddot{a}is\ddot{a}l\ddot{a}$ frequency.
\end{itemize}

Furthermore, our analysis identifies the conditions for the most unstable modes and provides a practical method for estimating their wavelength and growth rate. 
Applying these findings to a simplified CGM model with a linear temperature gradient, we derive the typical time and length scales associated with the formation of high-velocity clouds. 
Our theoretical estimation suggests that, for simulations discussing the formation fo high-velocity clouds (HVCs), a resolution of at least $\ll 7\ {\rm kpc}$ is required. 
Note that during the condensation, the Field length decreases monotonically.
The most severe condition is $\sim 10^{-5}$ kpc for the HVC's with $n = 10\ {\rm cm}^{-3}$ and $T=10^4\ {\rm K}$.
This is particularly relevant when considering observational facts: while the gas supply rate from the CGM is comparable to the star formation rate of the Milky Way, the observed mass and size of HVCs is smaller than our theoretical expectations (e.g., \citealp{Miller+2009, Nagata+2025}).

This discrepancy may be attributed to several causes. 
For example, while observations can only trace the neutral phase, recent simulations show that a significant amount of ionized gas envelops the neutral phase (\citealp{Lucchini+2025}). 
Thus, the total mass including both the ionized and neutral phases may be a few orders of magnitude larger than the quantities estimated from observations. 
\cite{Shimoda_etal2024} also pointed out the similar possibility that the gas with $T\lesssim 10^6\ {\rm K}$, which is bright at far ultraviolet band and observationally "dark" due to the attenuation, possibly dominates the mass of accretion flow.
Another possibility is the temperature structure of the CGM. 
Previous works suggest that the CGM has a multi-phase structure and turbulent mixing layers (\citealp{Wakker_etal2012, Tumlinson_etal2017, Putman_etal2012b, McCourt_etal2012, Ledos_etal2024, Ramesh_etal2024, Yao_etal2025}). 
Therefore, if the average temperature becomes locally smaller, the most unstable wavelength could become smaller than our estimation. 
Another concern is cloud survival (e.g., \citealp{Li+2020, Richie+2024}), specifically whether HVCs can truly reach the mid-plane of the Galactic disk after interacting with the relatively dense gas above the disk. 
Evidently, more detailed analysis of the non-linear dynamics of infalling clouds is needed to understand the actual consequences of thermal instability in the Galactic halo. 
High-resolution simulations will enable us to examine these possibilities, which constitutes one of our future works.

\section*{Acknowledgements}

This work was supported by JSPS KAKENHI, Grant Nos. 24KJ1302 and 25H00394.

\section*{Data Availability}

The data underlying this article will be shared on reasonable request to the corresponding author.
 



\bibliographystyle{mnras}
\bibliography{example} 






\bsp	
\label{lastpage}
\end{document}